%% file: Formatting-Instructions-LaTeX-2026.tex
\documentclass[letterpaper]{article} 
\usepackage{aaai2026}  
\usepackage{times}  
\usepackage{helvet}  
\usepackage{courier}  
\usepackage[hyphens]{url}  
\usepackage{graphicx} 
\usepackage{natbib}  
\usepackage{caption} 
\usepackage{algorithm}
\usepackage{booktabs}
\usepackage{subcaption}
\usepackage{multirow}
\usepackage{amsmath}
\usepackage{verbatim}
\usepackage{algpseudocode}
\usepackage{newfloat}
\usepackage{listings}
\DeclareCaptionStyle{ruled}{labelfont=normalfont,labelsep=colon,strut=off} 
\floatstyle{ruled}
\newfloat{listing}{tb}{lst}{}
\floatname{listing}{Listing}
\title{Enhancing All-to-X Backdoor Attacks with Optimized Target Class Mapping}
\author{
    Lei Wang\textsuperscript{\rm 1}, Yulong Tian\textsuperscript{\rm 1,\rm 2}\thanks{Yulong Tian is the corresponding author.}, Hao Han\textsuperscript{\rm 1},  
    Fengyuan Xu\textsuperscript{\rm 2}
}
\affiliations{
    \textsuperscript{\rm 1}College of Computer Science and Technology, Nanjing University of Aeronautics and Astronautics, China\\
    \textsuperscript{\rm 2}National Key Lab for Novel Software Technology, Nanjing University, China\\
    lei.wang@nuaa.edu.cn, yulong.tian@nuaa.edu.cn, hhan@nuaa.edu.cn, fengyuan.xu@nju.edu.cn
}

\usepackage{bibentry}

\begin{document}

\maketitle

\begin{abstract}
Backdoor attacks pose severe threats to machine learning systems, prompting extensive research in this area. However, most existing work focuses on single-target All-to-One (A2O) attacks, overlooking the more complex All-to-X (A2X) attacks with multiple target classes, which are often assumed to have low attack success rates. In this paper, we first demonstrate that A2X attacks are robust against state-of-the-art defenses. We then propose a novel attack strategy that enhances the success rate of A2X attacks while maintaining robustness by optimizing grouping and target class assignment mechanisms. Our method improves the attack success rate by up to 28\%, with average improvements of 6.7\%, 16.4\%, 14.1\% on CIFAR10, CIFAR100, and Tiny-ImageNet, respectively. We anticipate that this study will raise awareness of A2X attacks and stimulate further research in this under-explored area. Our code is available at \url{https://github.com/kazefjj/A2X-backdoor}.
\end{abstract}

\input{01_introduction}
\input{02_related_work}
\input{03_A2X}

\input{04_method}

\input{05_exper}

\input{06_conclusion}

\section{Acknowledgments}
This work was supported in part by the Natural Science Foundation of China (\#62402218 and \#62272224), the Natural Science Foundation of Jiangsu Province (\#BK20241378), the Yangtze River Delta Science and Technology Innovation Community Joint Research Project (\#2024CSJZN00400), the Postdoctoral Fellowship Program of CPSF (\#GZC20242229), and the Jiangsu Funding Program for Excellent Postdoctoral Talent. 

\bibliography{aaai2026}
\input{08_appendix}
\end{document}

%% file: 01_introduction.tex
\section{Introduction}
\label{sec:introduction}

Deep learning has achieved remarkable success across various domains, including face recognition~\cite{taigman2014deepface,schroff2015facenet}, autonomous driving~\cite{jin2022eigenlanes,zeng2022motr,jiang2023vad}, and security surveillance~\cite{ribeiro2018study,benfold2011stable}. Despite these advancements, its opaque nature also exposes deep learning systems to significant security threats. Among these, backdoor attacks pose one of the most severe risks. In such attacks, adversaries inject malicious functionalities into deep learning models, typically by inserting poisoned samples into the training dataset~\cite{gu2019badnets,chen2017targeted,barni2019new} or modifying model parameters~\cite{liu2018trojaning,tang2020embarrassingly}.
These injected hidden malicious behaviors remain dormant under normal inputs but activate when the input contains pre-defined triggers.

\begin{figure}[t]
\centering
\includegraphics[width=0.85\columnwidth]{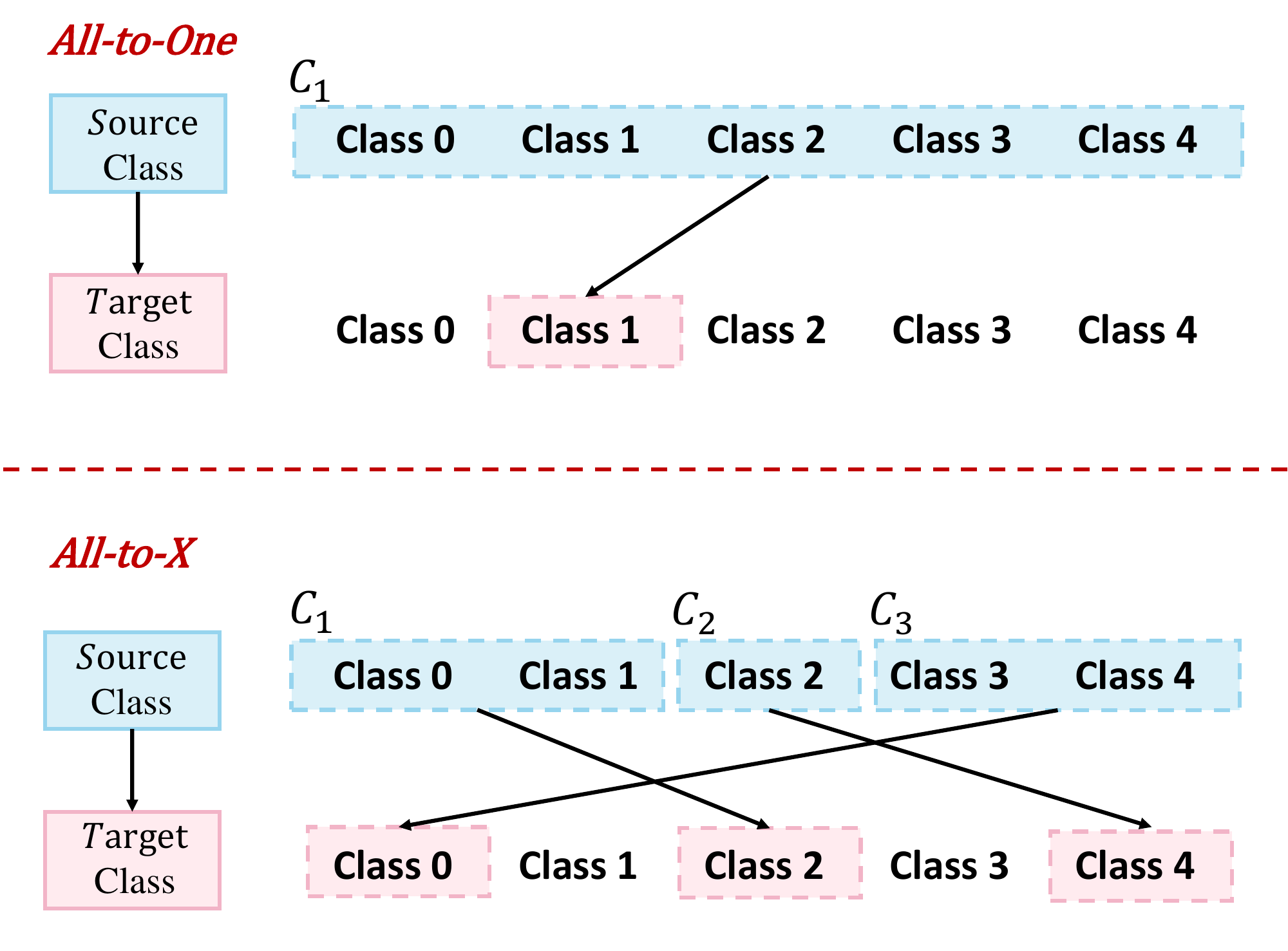} 
\caption{Comparison of A2O and A2X attacks. In A2O attacks, all triggered samples from source classes are misclassified into a single target class (Class 1). In A2X attacks, source classes are clustered into $X$ groups ($X$=3 shown here), with each group assigned a distinct target class. Triggered samples from each group are then 
 misclassified to their group's designated target class.}
\label{fig:a2x_illustration}
\end{figure}
Since backdoors can lead to severe consequences in real-world applications, various defensive methods have been proposed~\cite{li2021anti,wang2019neural,guo2023scale,hou2024ibd,gao2019strip,huang2022backdoor}. However, existing defensive research has predominantly focused on All-to-One (A2O)~\cite{gu2019badnets,chen2017targeted} attacks, where all poisoned samples are misclassified into a single target class, while largely overlooking All-to-X (A2X)~\cite{gu2019badnets,doan2021lira,cai2024towards} attacks, which distribute misclassifications across multiple target classes (Figure~\ref{fig:a2x_illustration} illustrates the differences between A2O and A2X attacks). This oversight is critical because A2X attacks inherently exhibit both enhanced robustness against defensive methods. Their distributed misclassification behaviors better mimic natural model errors, making them harder to detect. Despite this advantage, A2X attacks remain understudied due to a key practical limitation: their multi-target nature increases task complexity, often leading to unsatisfactory attack success rates.

Therefore, in this paper, we investigate how to enhance the effectiveness of A2X attacks and highlight their underestimated risks. Through an analysis of prior work, we find that existing A2X attacks employ overly simplistic strategies for target class mapping, such as misclassifying the $i$-th class as the $(i+1)$-th class \cite{gu2019badnets,doan2021lira,cai2024towards} or using random mappings \cite{gu2019badnets,li2022backdoor,nguyen2020input}. These strategies ignore the significant impact of mapping selection on attack performance. In fact, our experimental results demonstrate that the attack effectiveness of A2X attacks can be greatly improved by carefully designing mapping strategies.

To overcome the limitations in attack effectiveness, we propose a systematic two-step approach for optimizing target class mappings in A2X attacks. A2X attacks operate by (1) recognizing source class groups and (2) predicting triggered samples from each group into their associated target classes (see Figure~\ref{fig:a2x_illustration}). Our method optimizes these two steps independently.
First, we group semantically similar source classes by leveraging feature representations from a surrogate model, measuring class similarity through clustering in the embedding space. This grouping strategy enhances inter-group distinction, significantly simplifying source class recognition.
Next, we frame target class assignment as an optimization problem, maximizing the feature-space distance between source clusters and their assigned target classes via bipartite graph matching. This approach minimizes feature interference between source and target classes, thereby easing the model's learning burden.

Our main contributions are summarized as follows:
\begin{itemize}
\item We reveal that state-of-the-art backdoor defenses exhibit insufficient performance against A2X attacks, as their assumptions primarily hold for A2O attacks.

\item We design novel target class mapping strategies and demonstrate that, contrary to conventional belief, A2X attacks can achieve high attack success rates. To the best of our knowledge, our work is the first to systematically enhance the effectiveness of A2X attacks.
 
\item We validate the effectiveness of our proposed A2X attacks through extensive experiments. Our design significantly improves the attack success rate compared to existing methods while demonstrating minimal dependence on attacker knowledge and high transferability.

\end{itemize}

%% file: 02_related_work.tex
\section{Related Work}

Backdoor attacks can be categorized as All-to-One (A2O) or All-to-X (A2X) based on their target class selection. While most research focuses on A2O attacks that map all triggered samples to a single target class~\cite{gu2019badnets,chen2017targeted,barni2019new,li2021invisible}, A2X attacks employ multiple target classes~\cite{gu2019badnets,doan2021lira,li2022backdoor,nguyen2020input,cai2024towards}. Notable examples include the All-to-All variant with cyclic class mappings \cite{gu2019badnets,doan2021lira,cai2024towards} and random one-to-one source-target assignments \cite{gu2019badnets,li2022backdoor,nguyen2020input}. These A2X attacks have received less attention due to their low attack success rates. To address this gap, in this paper, we propose a novel mapping selection method to enhance the attack success rate of A2X backdoor attacks.

Researchers have designed various approaches to improve attack robustness against defensive measures (see backdoor defenses in Appendix~C) for A2O attacks, including developing more invisible triggers~\cite{chen2017targeted,barni2019new,li2021invisible}, 
designing conditional backdoors~\cite{tian2022stealthy,dong2023mind,duan2024conditional}, 
and selecting more effective poisoned samples~\cite{xia2022data,wu2023computation,gao2023not}. 
However, as we show in Section~\ref{sec:good_bad}, A2X attacks offer orthogonal robustness benefits while facing distinct challenges in attack effectiveness. 

%% file: 03_A2X.tex
\section{The Good and Bad of A2X Attacks}
\label{sec:good_bad}

\begin{table*}[htbp]
\centering
\fontsize{9}{9}\selectfont
\setlength{\tabcolsep}{2pt}
\begin{tabular}{c|c|c|c|c|c|c|c}
\toprule
\textbf{Defenses}$\rightarrow$ & {\textbf{No defense}} & \textbf{ABL} & \textbf{V\&B} &\textbf{IBD} &\textbf{SCALE} & \textbf{NC} & \textbf{FP} \\
\textbf{Value of X}$\downarrow$ & ASR  & ASR  & ASR & ASR  & ASR   & ASR  & ASR  \\
\midrule
\textbf{X=1 (A2O)}  & 97.4\scalebox{0.8}{$\pm$0.1}  & \textbf{2.0}\scalebox{0.8}{$\pm$1.4}   & 
\textbf{0.7}\scalebox{0.8}{$\pm$0.6}   & 
\textbf{0.6}\scalebox{0.8}{$\pm$0.6} 
& \textbf{12.1}\scalebox{0.8}{$\pm$10.2} 
& \textbf{6.0}\scalebox{0.8}{$\pm$9.2} 
& \textbf{1.3}\scalebox{0.8}{$\pm$0.7}  \\

\textbf{X=2}  & 89.2\scalebox{0.8}{$\pm$1.5}  & 86.0\scalebox{0.8}{$\pm$4.1} &   
\textbf{4.6}\scalebox{0.8}{$\pm$5.7} & 
\textbf{5.9}\scalebox{0.8}{$\pm$8.6} & 
46.6\scalebox{0.8}{$\pm$12.0}  & 90.3\scalebox{0.8}{$\pm$0.6}  & 79.3\scalebox{0.8}{$\pm$17.6}  \\

\textbf{X=5}   & 87.6\scalebox{0.8}{$\pm$1.2}  & 78.4\scalebox{0.8}{$\pm$5.3}   & 
58.9\scalebox{0.8}{$\pm$13.2} & 
43.0\scalebox{0.8}{$\pm$35.6} &  53.9\scalebox{0.8}{$\pm$12.0}  & 86.0\scalebox{0.8}{$\pm$2.2}  & 71.5\scalebox{0.8}{$\pm$18.5} \\

\textbf{X=8}   & 87.4\scalebox{0.8}{$\pm$0.6}  & 81.2\scalebox{0.8}{$\pm$3.0}   
& 85.7\scalebox{0.8}{$\pm$6.0} & 51.6\scalebox{0.8}{$\pm$34.0}  & 68.8\scalebox{0.8}{$\pm$4.2}  & 87.8\scalebox{0.8}{$\pm$0.5}  & 47.1\scalebox{0.8}{$\pm$40.4}   \\

\textbf{X=10}  & 86.7\scalebox{0.8}{$\pm$0.5}  & 79.0\scalebox{0.8}{$\pm$3.7}  
& 85.4\scalebox{0.8}{$\pm$1.5} 
& 72.5\scalebox{0.8}{$\pm$9.1}  
& 73.3\scalebox{0.8}{$\pm$4.6} 
& 86.4\scalebox{0.8}{$\pm$1.5}  &\textbf{5.8}\scalebox{0.8}{$\pm$10.1}  \\

\bottomrule
\end{tabular}
\caption{Performance of Defense Methods Against existing A2X Attacks  on CIFAR10 with ResNet18. Results are reported as \{mean\} $\pm$ \{standard deviation \} of five repeated trials. ASR(\%) denotes Attack Success Rate. X represents the number of target classes. Bold values indicate results lower than 20\%.
}
\label{tab:defense_comparison}
\end{table*}

\subsection{Definition of A2X Attacks} \label{sec:def_A2X}

While the commonly studied A2O attacks classify all samples with specific triggers into only one target class for each model, A2X attacks redirect those samples into multiple target classes ($X$ \textgreater 1).

Specifically, given input sample $x$ and its corresponding ground-truth label $y \in \mathcal{Y} = {0,...,K-1}$, where $K$ is the number of all possible labels, the deep learning model with a backdoor predict a target class for triggered version of input $x_{tr}$, $tr$ denotes the attacker-specified trigger, and in the A2X attack, the trigger remains identical for all target classes. The determination of the target class only depends on the function $\mathcal{G}(y)$, where $\mathcal{G}$ maps source classes into target class(es).

Unlike A2O attacks that only have one target class ($|\{\mathcal{G}(y)\}| = 1$), A2X attacks misclassify triggered inputs into X target classes ($|\{\mathcal{G}(y)\}| = X$). In an A2X attack, 
when determining the class mapping, all source classes are divided into X groups, each associated with a different target class. The lower part of Figure~\ref{fig:a2x_illustration} illustrates an example of $X=3$, where the source classes are categorized into three groups ($C_1$, $C_2$, and $C_3$) and are mapped to target classes 2, 4, and 0, respectively. When the number of target classes and source classes is the same ($X = K$), the attack is reduced to All-to-All attacks. A typical class mapping method~\cite{gu2019badnets,nguyen2021wanet,doan2021lira} for this case is $\mathcal{G}(y) = (y + 1) \mod K$, meaning each class is misclassified as the next one in a cyclic manner.

We note that the One-to-N (O2N) attack~\cite{xue2020one} also supports multi-target settings. However, O2N employs multiple distinct triggers to target different classes, essentially acting as a combination of several A2O attacks. Consequently, some existing defenses designed for A2O attacks can mitigate O2N attacks. In contrast, our proposed A2X attack utilizes only a single trigger and demonstrates inherent robustness against current defenses. Additional details regarding this comparison are provided in Appendix~B.

\subsection{Robustness against Defenses of A2X Attacks}

\label{defense setting}

As most existing defenses are primarily designed for A2O attacks, it is unclear whether they remain effective for A2X attacks. In this section, we show existing representative SOTA 
backdoor defenses are ineffective to A2X attacks. 

We construct the A2X attack using ResNet18 models with the CIFAR10 dataset. We utilizing a 3$\times$3 white square as the trigger pattern and choose a poisoning rate of 5\%, following the setup of BadNets~\cite{gu2019badnets}. The value of number of target classes $X$ is set to 1, 2, 5, 8, 10, respectively. When mapping the source classes to target classes ($\mathcal{G}(\cdot)$), following~\cite{gu2019badnets,doan2021lira,cai2024towards}, we employ $\mathcal{G}(y) = (y + 1) \mod K$ for $X$=10 and random mapping for other $X$ values.

We consider six representive backdoor defenses. Anti-Backdoor Learning (ABL)~\cite{li2021anti}, the Victim and the Beneficiary (V\&B)~\cite{zhu2023victim}, FinePruning(FP)~\cite{liu2018fine}, Input-level Backdoor Detection (IBD)~\cite{hou2024ibd}, SCALE-UP(SCALE )~\cite{guo2023scale}, Neural Cleanse (NC)~\cite{wang2019neural}. These defenses are implemented using the open-source BackdoorBox toolkit~\cite{li2023backdoorbox}.
Since IBD and SCALE are originally designed for backdoor detection (not removal), we integrate the unlearning method proposed by \cite{li2021anti} to enable backdoor removal, ensuring fair comparison across all approaches. Detailed configurations are provided in Appendix~D.3.

\textbf{Experimental Results:} Table~\ref{tab:defense_comparison} reports our experimental results. The results demonstrate that existing defenses are highly effective against A2O attacks but show unsatisfactory performance against A2X attacks. \textbf{Attack Success Rate (ASR)} in the table is the proportion of poisoned samples misclassified into the target classes. After applying backdoor defense, the attack success rate of A2O attacks decreases from 97.4\% to less than 12.1\% (the first row of Table~\ref{tab:defense_comparison}), highlighting the effectiveness of defensive methods. In contrast, for A2X attack, none of the methods can remain effective (working only in a limited number of cases).
 Moreover, the attack success rate exceeds 70\% in half of the experimental settings.
The ineffectiveness of these defenses stems from the fact that their underlying assumptions do not hold for A2X attacks. For instance, the ABL method assumes that the model learns the backdoor task faster compared to the main model training task. This assumption is reasonable for A2O attacks, where predicting all samples with a specific trigger into a single target class is simpler than the main training objective.
However, the backdoor task becomes significantly more complex for A2X attacks, especially as the value of X increases. Consequently, the backdoor task is not necessarily easier than the main model training task, which renders the ABL defense ineffective. A detailed analysis of the ineffectiveness of those defenses is provided in Appendix~E.

\subsection{Effectiveness of A2X Attacks}
Although A2X attacks exhibit inherent robustness against existing defenses, the backdoor task of mapping source classes into different target classes is often too complex for the model to learn. This complexity can lead to low attack success rates, especially under low poisoning rates.

\begin{figure}[t]
\centering
\includegraphics[width=0.95\columnwidth]{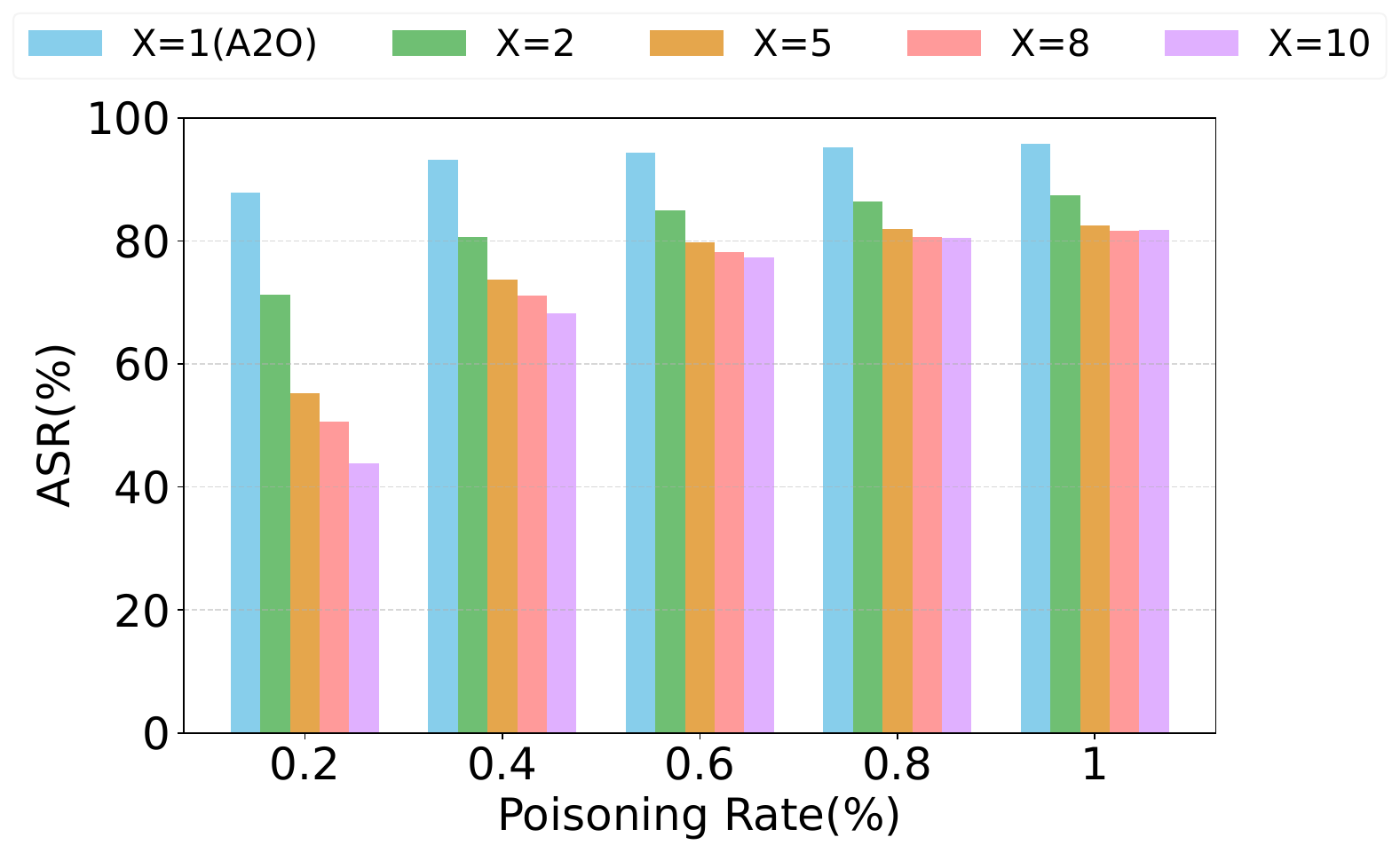} 
\caption{The Attack Success Rate of A2X Attacks under Different Poisoning Rates on CIFAR10 with ResNet18. 
}
\label{fig:low_attack_success_rate}
\end{figure}

Figure~\ref{fig:low_attack_success_rate} reports the attack success rates of existing A2X attacks on CIFAR10 with varying poisoning rates. The results demonstrate that A2X attacks consistently achieve lower attack success rates compared to A2O attacks (the case where $X=1$) across all settings, with attack success rates decreasing as $X$ increases. For example, when the poisoning rate is 0.2\%, the A2O attack achieves a high success rate of 87.8\%, while A2X attacks yield success rates below 72\%. This rate drops further to 50.7\% when $X$ increases to 5, and decreases to  43.8\% when $X=10$.

This low attack success rate explains why prior defensive methods tend to overlook A2X attacks. Therefore, in this paper, we highlight the potential harm of A2X attacks by designing new methods that significantly improve their attack success rate. Our experimental results in Section~\ref{sec:evaluation} demonstrate that our methods can increase the attack success rate by up to 28\%.

%% file: 04_method.tex
\section{Methodology}
\label{sec:method}

\subsection{Threat Model}
We consider the scenario where the adversary acts as the data provider and can inject poisoned samples into the training dataset. The victim trains a model using the poisoned dataset, and the adversary has no control over the training process. The adversary's goal is to inject a hidden backdoor into the victim’s model, which behaves normally on clean inputs but causes attacker-specified misclassifications when presented with inputs containing triggers. 

Following \cite{xia2022data,wu2023computation}, we consider two distinct scenarios based on the attacker's knowledge: (1) the adversary possesses prior knowledge of the victim's training details (including model architectures and optimizers), and (2) such knowledge is unavailable. As demonstrated in Section~\ref{sec:transferability}, our method achieves consistent effectiveness in both scenarios.

\subsection{Overview}
\label{motivation}

Our research goal is to improve the attack success rate of A2X attacks while maintaining their robustness. Through careful analysis of existing A2X attack methods, we discovered that these methods typically employ overly simplistic target class mapping strategies, such as random mapping or mapping one class to the next class in a cyclic manner. We hypothesize that these simple class mapping approaches are the root cause of the ineffectiveness of A2X attacks, and aim to enhance attack effectiveness by optimizing the mapping strategy (The impact of mapping strategy on attack flexibility is discussed in Appendix G).

Recall that A2X attacks predict triggered samples originally belonging to each class group into a specific target class (Figure~\ref{fig:a2x_illustration} and Section~\ref{sec:def_A2X}). The underlying mechanism of the backdoor involves two key steps: (1) recognizing the class groups and (2) identifying the target class assigned to each class group, both the class grouping and target class assignment  can be sub-optimal, presenting significant opportunities for improvement.
\begin{figure}[h]
\centering
\includegraphics[width=0.7\columnwidth]{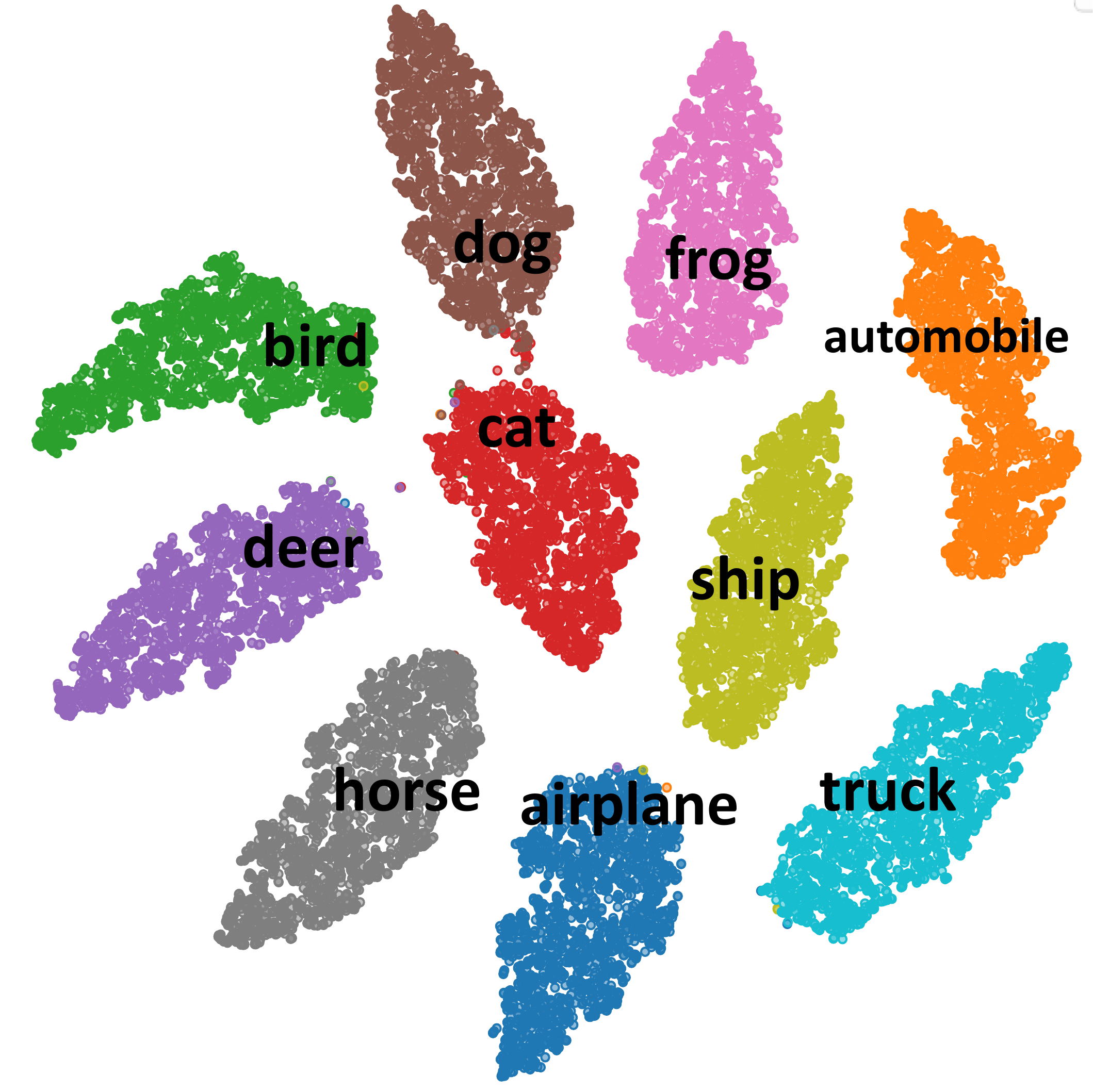} 
\caption{The t-SNE visualization of the CIFAR-10 dataset}
\label{fig:tsne-cifar10}
\end{figure}

\begin{figure}[t]
\centering
\includegraphics[width=0.85\columnwidth]{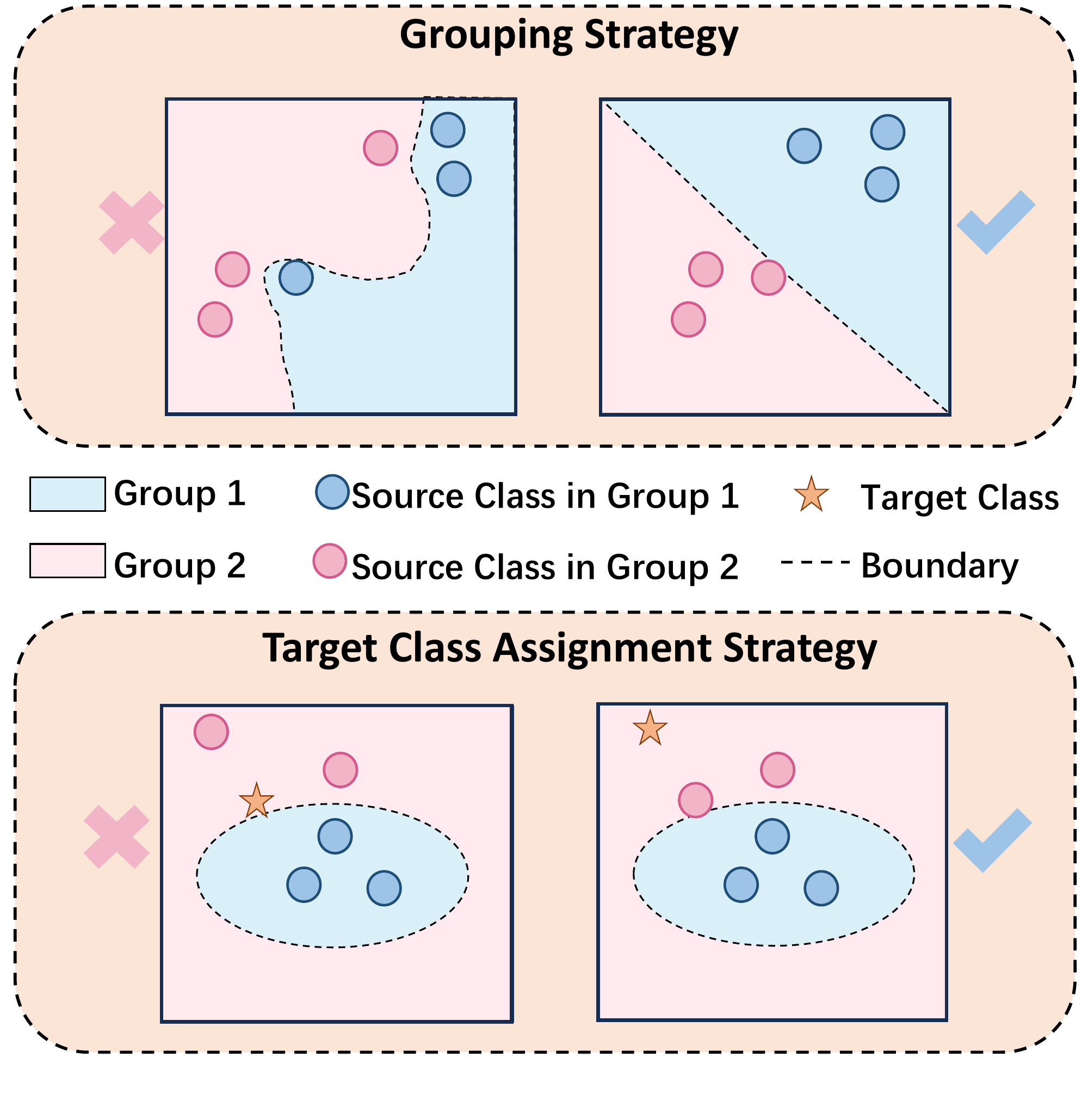} 
\caption{Comparison of our method with existing approaches. We first cluster similar classes into the same class groups, resulting in simpler decision boundaries that are easier to learn (upper part). We then select more distant target class for each class group to reduce feature interference during model training (lower part).}
\label{fig:design-comparison}
\end{figure}

For the class grouping aspect, it is evident that recognizing a class group containing similar classes (such as cat and dog in Figure~\ref{fig:tsne-cifar10}, which shows the t-SNE visualization of  CIFAR10) is simpler than recognizing a group consisting of dissimilar classes (such as dog and truck), as the former requires a less complex classification boundary (upper part of Figure~\ref{fig:design-comparison}). This observation suggests that we can design grouping strategies that avoid the sub-optimal results induced by random grouping and are easier for the model to learn, thereby achieving better attack effectiveness. \textit{As illustrated in the upper part of Figure~\ref{fig:design-comparison}, our solution for this aspect is to group similar classes together, forming a simpler classification boundary, thus making this task more learnable for the model (Section~\ref{sec:design_our_a2x}).}

For the target class assignment aspect, establishing backdoor mappings from source groups to target classes that are distant from the groups in the feature space is clearly easier than mapping to classes that are closer. For example, predicting triggered samples originally from classes cat and dog to the class truck is easier than predicting samples from classes cat and dog to the class bird (Figure~\ref{fig:tsne-cifar10}). When the feature distance between a source group and its corresponding target class is too small, feature overlap occurs, significantly interfering with the learning of the backdoor mapping and potentially resulting in a lower attack success rate. \textit{To address this challenge, as shown in the lower part of Figure~\ref{fig:design-comparison}, we propose a method that assigns target classes that are maximally distant from the source groups in the feature space to aviod feature confusion (Section~\ref{sec:design_our_a2x}).}

\subsection{The Design of Our A2X Attack} \label{sec:design_our_a2x}
In this section, we present the details of our method, including the similarity-based class grouping and distance-aware target class assignment. Algorithm~1 in Appendix~A presents the details of our design.


\textbf{Similarity-Based Class Grouping that Maximizes Intra-group Similarity.} Our class grouping method divides all source classes into $X$ groups, with the requirement that classes within each group exhibit high similarity. These grouped classes will later be used for target class assignment.

We first calculate the similarity between all classes in the dataset and then perform a clustering algorithm based on these similarities to produce $X$ clusters. We directly use these clusters as class groups. Specifically, we train a surrogate model $f_s$ on the original dataset and use this model to extract features for class-wise similarity calculation. Following existing practices \cite{paul2021deep,wu2023computation}, we calculate the position vector of class $i$ using the following equation:
\begin{equation} \label{eq:position_vec}
     P_i=\frac{1}{|\mathcal{X}^i|}\sum_{x^i \in\mathcal{X}^i} f_s(x^i,\theta)
\end{equation}
where $\mathcal{X}^i$ denotes all samples belonging to class $i$ in the original dataset, $\theta$ represents the parameters of the surrogate model, $x^i$ represents the sample from class $i$, and $f_s(x^i,\theta)$ indicates feature extraction from $x^i$ using the surrogate model $f_s$. The $x^i$ can be either a clean sample or a triggered sample. Empirically, we find that both choices yield similarly effective results; thus, we use the triggered sample as $x^i$ for our main experiments.
The position vector of each class is the average of features extracted from the samples of that class. Based on these position vectors, we measure the distance between class $i$ and class $j$ using the $\ell_2$ norm, which can be expressed as $d_{i,j} = \| P_i-P_j \|_2$. Discussions on different norm choices are in Appendix~F.7.

Considering that numerous high-performance clustering methods exist and that K-means already achieves satisfactory results, we simply adopt K-means for clustering the source classes, with the number of desired clusters set to $X$. The resulting class clusters are directly used as the class groups for our A2X attack.

\textbf{Distance-Aware Target Class Assignment that Maximizes the Distances Between Groups and Target Classes.} Our target class assignment method for backdoor mapping determines a target class for each of the $X$ class groups, while maximizing the distance between each group and its assigned target class. The distance between a group $C_i$ and its corresponding target class $\mathcal{G}(C_i)$ is defined as the sum of the distances from all classes within the group to the target class. Our objective can be formalized as:
\begin{equation} \label{eq:mapping_obj}
    \mathop{argmax}\limits_{\mathcal{G}} \sum_{C_i \in \mathcal{C}} \sum_{j \in C_i}d_{j,\mathcal{G}({C_i})} 
\end{equation}
where $\mathcal{C}$ denotes the set that includes all groups. This objective maximizes the sum of distances between each class group and its corresponding target class. This optimization problem can be reduced to a Maximum Bipartite Matching Problem and efficiently solved using existing methods. Specifically, we construct a bipartite graph using source class groups $\mathcal{C}={C_1,C_2,...,C_X}$ and all possible target classes $\mathcal{Y}={0,1,...,K-1}$ as nodes. The nodes within set $\mathcal{C}$ are not connected to each other, and similarly, nodes within set $\mathcal{Y}$ have no internal connections.
Edges exist only between nodes in $\mathcal{C}$ and nodes in $\mathcal{Y}$, resulting in a total of $K \times X$ edges. The weight of each edge is set to the distance between the corresponding source group and target class ($\sum_{j \in C_i}d_{j,\mathcal{G}(C_i)}$). Optimizing Equation~\ref{eq:mapping_obj} is equivalent to finding the Maximum Weight Bipartite Matching of this graph. In this paper, we employ the Hungarian Algorithm to efficiently solve this optimization problem.

%% file: 05_exper.tex
\section{Evaluation} \label{sec:evaluation}
\begin{figure*}[t]
\centering
\includegraphics[width=0.82\linewidth]{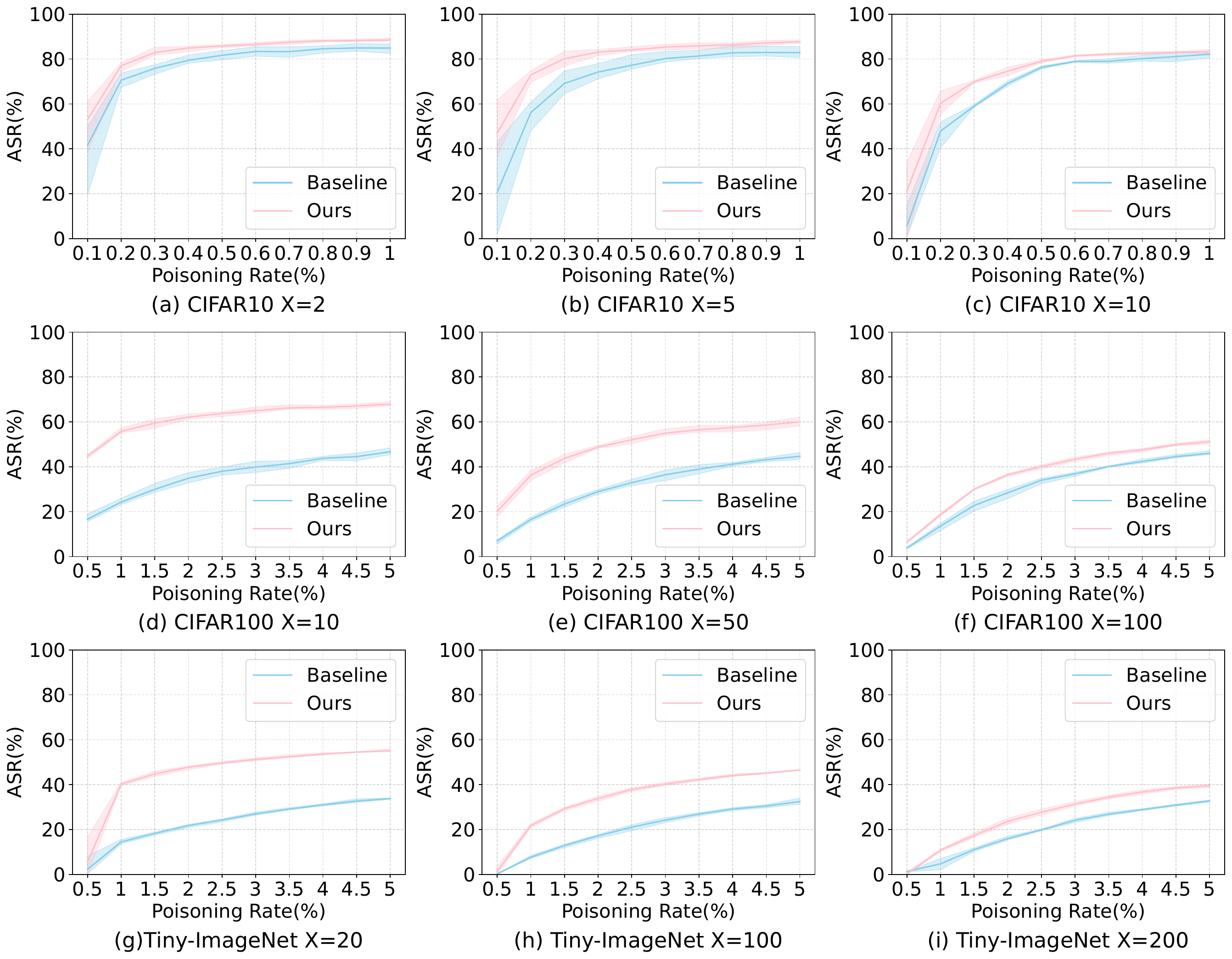}
\caption{Attack Success Rates across Different Poisoning Rates and Datasets. 
 ``Baseline'' lines are the results of the baseline methods, 
 and the ``Ours'' lines show the results of our proposed method. Lines represent average values from five repeated experiments, with shadow regions indicating standard deviations.}
\label{fig:main_results}

\end{figure*}
In this section, we systematically evaluate our proposed method. We first introduce the experimental settings, then present the main results, demonstrate the attack's effectiveness against defensive measures, examine its transferability across different victim model knowledge scenarios, and finally show the ablation results.

\subsection{Experimental Settings}

We conduct experiments on three commonly used datasets, including CIFAR10~\cite{krizhevsky2009learning}, CIFAR100~\cite{krizhevsky2009learning} and Tiny-ImageNet~\cite{deng2009imagenet}. We use three classicial model architectures for our experiments, including ResNet18~\cite{he2016deep}, VGG16~\cite{simonyan2014very} and MobileNetV2~\cite{howard2017mobilenets}. 

When training backdoor models, we follow the data poisoning methods of BadNets~\cite{gu2019badnets} and employ five representative types of backdoor triggers, including BadNets with white square pattern (BD-white)~\cite{gu2019badnets}, BadNets with random square pattern (BD-random)~\cite{gu2019badnets}, Label-Consistent attack (LC)~\cite{turner2019label}, Blend~\cite{chen2017targeted}, and Sinusoidal signal attack (SIG)~\cite{barni2019new}. We adopt BD-white as the default trigger throughout our experiments. As for the backdoor defense methods, we use the six defensive methods used in Section~\ref{defense setting}. For each experimental setting, we conduct five repeated experiments with different random seeds and report the average value with the standard deviation. More detailed experimental settings are in Appendix~D.

\subsection{Main Results of Our Proposed A2X Attack} \label{sec:main_results}

We evaluate our proposed A2X attack with three different values of target classes ($X$) and various poisoning rates using ResNet18. Following~\cite{gu2019badnets,doan2021lira,cai2024towards}, we employ cyclic class mapping $\mathcal{G}(y) = (y + 1) \mod K$ for $X$=10 and random mapping for other $X$ values as baseline, where all other settings remain the same except for the mapping.

\textbf{Experimental Results:}  Figure~\ref{fig:main_results} shows our results. Our method consistently achieves higher attack success rates compared to the baseline across all three datasets, with one exception noted below. For CIFAR10, the attack success rates are improved by an average of 5.2\%, 9.2\%, and 5.8\% for $X$ values of 2, 5, and 10, respectively. For CIFAR100, the attack success rates are improved by 25.8\%, 17.5\%, and 5.7\% on average for $X$ values of 10, 50, and 100, respectively. For Tiny-ImageNet, when the poisoning rate is 0.5\%, neither the baseline attack nor our proposed attack successfully learns the backdoor task due to the extremely low poisoning rate. For other poisoning rates, the attack success rates show average improvements of 22.0\%, 14.0\%, and 6.4\% for $X$ values of 20, 100, and 200, respectively. The results also show that backdoored models maintain clean accuracy nearly identical to clean models (differences within 0.5\%), details can be found in Appendix~F.4.

From these results, we observe that the improvements brought by our proposed method are particularly notable for CIFAR100 and Tiny-ImageNet. This suggests the effectiveness of our class grouping and target class assignment methods in complex scenarios. When there are only a few source classes (CIFAR10 only has 10 source classes), the simple mapping methods used by existing approaches can sometimes find relatively optimal mappings thus achieving relative high attack success rates. However, in cases with large number of source classes (100 for CIFAR100 and 200 for Tiny-ImageNet), simple methods have little chance of finding optimal mappings. Our proposed method effectively identifies optimal class groupings and  target class assignments, thus achieving superior results. Results with other triggers are provided in Appendix~F.3.

\begin{table*}[t]

\centering
\setlength{\tabcolsep}{2pt}
\fontsize{9}{9}\selectfont
\begin{tabular}{ll|cc|cc|cc|cc}
\toprule
 
\multicolumn{2}{c|}{\textbf{Value of X}$\rightarrow$}  
& \multicolumn{2}{c|}{\textbf{X=2}} 
& \multicolumn{2}{c|}{\textbf{X=5}} 
& \multicolumn{2}{c|}{\textbf{X=8}} 
& \multicolumn{2}{c}{\textbf{X=10}} \\
\textbf{Rate}$\downarrow$ &\textbf{Method}$\downarrow$ & baseline & \textbf{ours} & baseline & \textbf{ours} & baseline & \textbf{ours} & baseline & \textbf{ours} \\
\midrule
\multirow{6}{*}{\textbf{0.2\%}} 

& \textbf{NC}   & 34.8\scalebox{0.8}{$\pm$31.5} & \textbf{41.3}\scalebox{0.8}{$\pm$32.5} & 32.0\scalebox{0.8}{$\pm$25.5} & \textbf{62.0}\scalebox{0.8}{$\pm$15.2} & 44.7\scalebox{0.8}{$\pm$8.7} & \textbf{47.8}\scalebox{0.8}{$\pm$23.7}          & 40.9\scalebox{0.8}{$\pm$5.3} & \textbf{55.5}\scalebox{0.8}{$\pm$14.7} \\
& \textbf{IBD}  & \textbf{26.0}\scalebox{0.8}{$\pm$12.1} & 21.6\scalebox{0.8}{$\pm$17.6} & \textbf{45.7}\scalebox{0.8}{$\pm$7.5} & 39.1\scalebox{0.8}{$\pm$16.5}  & 44.6\scalebox{0.8}{$\pm$6.9} & \textbf{46.6}\scalebox{0.8}{$\pm$4.4}  & 35.0\scalebox{0.8}{$\pm$10.7} & \textbf{45.7}\scalebox{0.8}{$\pm$5.4} \\

& \textbf{SCALE}  & \textbf{68.3}\scalebox{0.8}{$\pm$7.4} & 47.2\scalebox{0.8}{$\pm$19.5} & 52.2\scalebox{0.8}{$\pm$10.3} & \textbf{62.2}\scalebox{0.8}{$\pm$9.7}  & 45.3\scalebox{0.8}{$\pm$6.4} & \textbf{48.8}\scalebox{0.8}{$\pm$8.1}  & 37.9\scalebox{0.8}{$\pm$3.2} & \textbf{52.8}\scalebox{0.8}{$\pm$9.6} \\

& \textbf{ABL} & 3.6\scalebox{0.8}{$\pm$0.6}  & \textbf{3.9}\scalebox{0.8}{$\pm$1.8} & \textbf{2.4}\scalebox{0.8}{$\pm$0.5}  & 0.8\scalebox{0.8}{$\pm$0.7} & \textbf{2.3}\scalebox{0.8}{$\pm$1.2}  & 0.9\scalebox{0.8}{$\pm$0.4}          & \textbf{2.3}\scalebox{0.8}{$\pm$0.7}  & 0.7\scalebox{0.8}{$\pm$0.4} \\

& \textbf{V\&B} & 49.3\scalebox{0.8}{$\pm$8.8}  & \textbf{51.2}\scalebox{0.8}{$\pm$23.8} & 41.4\scalebox{0.8}{$\pm$20.5}  & \textbf{67.9}\scalebox{0.8}{$\pm$12.6} &42.8\scalebox{0.8}{$\pm$6.1}  &  \textbf{67.8}\scalebox{0.8}{$\pm$6.2}  & 13.6\scalebox{0.8}{$\pm$21.0}  & \textbf{56.7}\scalebox{0.8}{$\pm$20.1} \\

& \textbf{FP}  &  \textbf{50.4}\scalebox{0.8}{$\pm$24.1}  &49.1\scalebox{0.8}{$\pm$28.9} & 30.4\scalebox{0.8}{$\pm$8.7}  & \textbf{33.2}\scalebox{0.8}{$\pm$28.0} & 20.9\scalebox{0.8}{$\pm$19.4}  & \textbf{47.4}\scalebox{0.8}{$\pm$9.8}          & 23.1\scalebox{0.8}{$\pm$11.5}  & \textbf{47.4}\scalebox{0.8}{$\pm$15.5} \\

\midrule
\multirow{6}{*}{\textbf{0.4\%}} 

& \textbf{NC}  & 45.4\scalebox{0.8}{$\pm$35.9} & \textbf{74.6}\scalebox{0.8}{$\pm$33.7} & 60.6\scalebox{0.8}{$\pm$25.3} & \textbf{65.7}\scalebox{0.8}{$\pm$10.7} & \textbf{67.6}\scalebox{0.8}{$\pm$10.8} &  64.7\scalebox{0.8}{$\pm$16.2} & 43.4\scalebox{0.8}{$\pm$21.7} & \textbf{66.8}\scalebox{0.8}{$\pm$14.5} \\
& \textbf{IBD}   & 9.7\scalebox{0.8}{$\pm$6.4} & \textbf{23.6}\scalebox{0.8}{$\pm$29.0} & 33.9\scalebox{0.8}{$\pm$15.8} & \textbf{48.8}\scalebox{0.8}{$\pm$22.4}  & 44.5\scalebox{0.8}{$\pm$11.8} & \textbf{57.3}\scalebox{0.8}{$\pm$8.5} & 53.7\scalebox{0.8}{$\pm$3.8} & \textbf{57.5}\scalebox{0.8}{$\pm$5.2} \\

& \textbf{SCALE} &77.2\scalebox{0.8}{$\pm$5.6} &  \textbf{82.2}\scalebox{0.8}{$\pm$2.1} & 70.5\scalebox{0.8}{$\pm$1.7} & \textbf{80.9}\scalebox{0.8}{$\pm$1.7}  & 63.5\scalebox{0.8}{$\pm$8.0} & \textbf{70.2}\scalebox{0.8}{$\pm$3.1}  & 51.5\scalebox{0.8}{$\pm$4.6} & \textbf{65.8}\scalebox{0.8}{$\pm$4.6} \\

& \textbf{ABL}  & 34.8\scalebox{0.8}{$\pm$15.4}  & \textbf{51.4}\scalebox{0.8}{$\pm$11.1}         & 8.2\scalebox{0.8}{$\pm$4.1}  & \textbf{43.7}\scalebox{0.8}{$\pm$7.7}          & 8.4\scalebox{0.8}{$\pm$3.0}  & \textbf{19.0}\scalebox{0.8}{$\pm$3.7}          & 10.3\scalebox{0.8}{$\pm$7.0}  & \textbf{22.4}\scalebox{0.8}{$\pm$6.4} \\

& \textbf{V\&B}  & 62.7\scalebox{0.8}{$\pm$5.0}  & \textbf{66.4}\scalebox{0.8}{$\pm$7.7} & 80.8\scalebox{0.8}{$\pm$1.7}  & \textbf{83.5}\scalebox{0.8}{$\pm$10.6} & 75.3\scalebox{0.8}{$\pm$7.6}  & \textbf{84.4}\scalebox{0.8}{$\pm$2.8}          & 75.3\scalebox{0.8}{$\pm$3.7}  & \textbf{85.0}\scalebox{0.8}{$\pm$1.6} \\

& \textbf{FP} & \textbf{63.2}\scalebox{0.8}{$\pm$19.9}  & 39.0\scalebox{0.8}{$\pm$30.0} & 51.0\scalebox{0.8}{$\pm$18.6}  & \textbf{60.8}\scalebox{0.8}{$\pm$30.1} & 37.5\scalebox{0.8}{$\pm$30.0}  & \textbf{72.2}\scalebox{0.8}{$\pm$2.2}          & 29.6\scalebox{0.8}{$\pm$25.4}  & \textbf{60.0}\scalebox{0.8}{$\pm$9.0} \\

\midrule
\multirow{6}{*}{\textbf{0.6\%}} 

& \textbf{NC}    & \textbf{30.3}\scalebox{0.8}{$\pm$29.9} & 26.5\scalebox{0.8}{$\pm$27.3} & 70.4\scalebox{0.8}{$\pm$16.8} & \textbf{79.4}\scalebox{0.8}{$\pm$7.4} & 73.8\scalebox{0.8}{$\pm$6.7} & \textbf{81.0}\scalebox{0.8}{$\pm$1.0}          & 68.1\scalebox{0.8}{$\pm$10.8} & \textbf{72.5}\scalebox{0.8}{$\pm$14.1} \\
& \textbf{IBD}  & 17.1\scalebox{0.8}{$\pm$12.6} & \textbf{18.8}\scalebox{0.8}{$\pm$31.8} & \textbf{60.9}\scalebox{0.8}{$\pm$15.3} & 41.1\scalebox{0.8}{$\pm$24.3}  & 57.4\scalebox{0.8}{$\pm$9.3} & \textbf{66.0}\scalebox{0.8}{$\pm$6.6}         & 63.9\scalebox{0.8}{$\pm$1.6} & \textbf{67.5}\scalebox{0.8}{$\pm$5.2} \\

& \textbf{SCALE}  & 70.6\scalebox{0.8}{$\pm$9.9} & \textbf{85.1}\scalebox{0.8}{$\pm$0.3} & 71.6\scalebox{0.8}{$\pm$3.3} & \textbf{81.3}\scalebox{0.8}{$\pm$13.7}  & 69.9\scalebox{0.8}{$\pm$5.4} & \textbf{74.7}\scalebox{0.8}{$\pm$1.6}  & 57.4\scalebox{0.8}{$\pm$2.1} & \textbf{66.5}\scalebox{0.8}{$\pm$2.9} \\

& \textbf{ABL}  & 68.5\scalebox{0.8}{$\pm$5.0}  &  \textbf{77.1}\scalebox{0.8}{$\pm$5.2}    & 51.2\scalebox{0.8}{$\pm$8.2}  & \textbf{72.3}\scalebox{0.8}{$\pm$3.0} & 45.6\scalebox{0.8}{$\pm$13.7}  & \textbf{62.6}\scalebox{0.8}{$\pm$6.4}          & 44.1\scalebox{0.8}{$\pm$6.4}  & \textbf{61.7}\scalebox{0.8}{$\pm$8.2} \\

& \textbf{V\&B}  & \textbf{70.0}\scalebox{0.8}{$\pm$8.8}  & 37.5\scalebox{0.8}{$\pm$10.3} & 80.8\scalebox{0.8}{$\pm$2.8}  & \textbf{88.1}\scalebox{0.8}{$\pm$2.1} & 83.7\scalebox{0.8}{$\pm$3.9}  & \textbf{88.0}\scalebox{0.8}{$\pm$4.2}          & 84.1\scalebox{0.8}{$\pm$3.4}  & \textbf{84.8}\scalebox{0.8}{$\pm$3.6} \\

& \textbf{FP}  & \textbf{55.6}\scalebox{0.8}{$\pm$18.8}  & 42.5\scalebox{0.8}{$\pm$34.6} & \textbf{70.4}\scalebox{0.8}{$\pm$1.7}  & 62.6\scalebox{0.8}{$\pm$28.3} & 63.8\scalebox{0.8}{$\pm$6.3}  & \textbf{64.9}\scalebox{0.8}{$\pm$30.0}          & 44.6\scalebox{0.8}{$\pm$18.4}  & \textbf{50.9}\scalebox{0.8}{$\pm$25.8} \\
\bottomrule
\end{tabular}
\caption{The Attack Success Rate(\%) of Our A2X Attack under Six Defensive Methods on CIFAR10 with ResNet18. Bolded values indicate higher ASR results under identical experimental configurations.}
\label{tab:results_under_defense}
\end{table*}

\subsection{Results of Our A2X Attack Under Defenses}
In this section , we demonstrate that our A2X attack achieves higher attack success rates compared to existing A2X attacks, even when defensive measures are employed. We conduct experiments on CIFAR10 with four different values of $X$. We select CIFAR10 for efficiency considerations---the defensive method NC works efficiently on CIFAR10 but requires excessive computational resources for the other two datasets that have over 100 classes. 

\textbf{Experimental Results:} The results are shown in Table~\ref{tab:results_under_defense}. Our method significantly enhances the attack success rates while preserving the robustness against defenses of the A2X attack. Our approach demonstrates superior attack success rates in 58 out of 72 experimental settings. The attack success rate improvements range from 0.3\% to 43.1\%, with average improvements of 5.7\%, 10.8\%, 9.8\% and 13.5\% for settings where $X$ values are 2, 5, 8, and 10, respectively. We note that when $X=2$, large standard deviations occasionally occur because some defenses occasionally succeed.

While our method achieves notable improvements in most settings, it shows slightly inferior attack success rates in some cases (14 out of 72). We observe that this occurs primarily (7 out of 14 settings) when the poisoning rate is extremely low (0.2\%). A possible explanation is that the backdoor task becomes too difficult to learn under such low poisoning rates, causing the attack to become unstable and resulting in relatively lower attack success rates for our method. Nevertheless, our attack remains effective even in these challenging conditions, still achieving better attack success rates in most experimental settings (17 out of 24).

\subsection{Attack Transferability of Our A2X Attack} \label{sec:transferability}

\begin{table}[t]
\centering
\setlength{\tabcolsep}{2pt}
\fontsize{9}{9}\selectfont
\begin{tabular}{cc|cccccc}
\toprule
 \multicolumn{2}{c|}{\textbf{Rate}$\downarrow$}   & \multicolumn{3}{c|}{\textbf{SGD}} & \multicolumn{3}{c}{\textbf{ADAM}} \\

& & \textbf{R(Same)} & \textbf{V} & \textbf{M} & \textbf{R(Same)} & \textbf{V} & \textbf{M}  \\
\midrule
\multicolumn{2}{c|}{\textbf{0.2\%}} & 73.0\scalebox{0.8}{$\pm$2.0}
& 72.9\scalebox{0.8}{$\pm$1.3} & 71.6\scalebox{0.8}{$\pm$4.5}
& 72.3\scalebox{0.8}{$\pm$4.1} & 69.7\scalebox{0.8}{$\pm$1.1}
& 74.5\scalebox{0.8}{$\pm$1.2} 
\\
\midrule
\multicolumn{2}{c|}{\textbf{0.4\%}} & 83.1\scalebox{0.8}{$\pm$1.1}
& 82.6\scalebox{0.8}{$\pm$1.0} & 82.7\scalebox{0.8}{$\pm$0.8}
& 82.2\scalebox{0.8}{$\pm$2.0} & 82.3\scalebox{0.8}{$\pm$1.9}
& 83.3\scalebox{0.8}{$\pm$1.6} 
\\
\midrule
\multicolumn{2}{c|}{\textbf{0.6\%}} & 85.3\scalebox{0.8}{$\pm$1.0}
& 84.3\scalebox{0.8}{$\pm$1.6} & 85.4\scalebox{0.8}{$\pm$0.5}
& 84.8\scalebox{0.8}{$\pm$0.4} & 84.1\scalebox{0.8}{$\pm$0.2}
& 85.0\scalebox{0.8}{$\pm$1.3} 
\\
\midrule
\multicolumn{2}{c|}{\textbf{0.8\%}} & 86.4\scalebox{0.8}{$\pm$1.0}
& 86.3\scalebox{0.8}{$\pm$0.9} & 86.7\scalebox{0.8}{$\pm$0.7}
& 86.2\scalebox{0.8}{$\pm$0.7} & 85.4\scalebox{0.8}{$\pm$0.3}
& 86.4\scalebox{0.8}{$\pm$0.7} 
\\
\midrule
\multicolumn{2}{c|}{\textbf{1\%}} & 87.7\scalebox{0.8}{$\pm$0.6}
&86.8\scalebox{0.8}{$\pm$0.9}  & 86.9\scalebox{0.8}{$\pm$0.5}
&87.2\scalebox{0.8}{$\pm$0.6}  & 86.6\scalebox{0.8}{$\pm$0.7}
& 86.2\scalebox{0.8}{$\pm$1.3}  
\\

\bottomrule
\end{tabular}
\caption{The Attack Success Rate (\%) of Our A2X Attack under Different Surrogate Model Training Configurations on CIFAR10. 
The ``R (Same)'' column are the settings where both the surrogate and victim models share identical training configurations. R, V and M denote ResNet18, VGG16 MobileNetV2, respectively.}
\label{tab:blackbox}

\end{table}

In our main experiments, the surrogate models were trained using the same configurations as those used in training the victim model. However, in real-world scenarios, attackers may have limited or no prior knowledge about the training configurations of the victim model, including model architecture and optimizer selection. For example, when the attacker is a data provider, they typically have no insight into the subsequent model training process.

We conduct experiments to demonstrate that our proposed attack does not depend on such knowledge, and attackers can still effectively launch attacks even when the surrogate model is trained using completely different configurations. We reused the attack setups for CIFAR10 with $X$=5 from Section~\ref{sec:main_results} as the ``same configuration'' baseline. For our transferability experiments, we design ``different configuration'' setttings by varying the surrogate model architectures (ResNet18, VGG16, and MobileNetV2) and optimizers (Adam and SGD), while keeping the victim model identical to that of the ``same configuration'' setting.

\textbf{Experimental Results:} Table~\ref{tab:blackbox} presents the results. ``R (Same)'' column in Table~\ref{tab:blackbox} shows the results of the ``same configuration'' experiments, while the other columns present the results of the ``different configuration'' experiments.  The results show that the two configurations achieve similarly effective results. 
Across all poisoning rate settings, the differences between the results from the ``same configuration'' and new ``different configuration'' scenarios are  within 3.28\% , and within 1\% for the majority of setings (18 out of 25 configurations). These results highlight the high transferability of our proposed methods and the minimal requirements regarding knowledge of the victim model.

\subsection{Ablation Studies}

Our proposed method comprises two key components: similarity-based class grouping and distance-aware target class assignment. To evaluate the contribution of each component to attack effectiveness, we conducted comprehensive ablation studies.

We design two intermediate configurations that isolate the impact of individual components, comparing them against both the baseline method and our complete approach. The results in Appendix~F.5 demonstrate that while both intermediate configurations achieve lower attack success rates than our full method, they consistently outperform the baseline approach that relies on random strategies, highlighting the effectiveness of each component of our design.

%% file: 06_conclusion.tex
\section{Conclusion}

In this paper, we designed a similarity-based class grouping method and a distance-aware target class assignment approach to replace the overly simplistic strategies used in existing A2X attacks. Our experimental results demonstrate that our proposed methods significantly enhance the attack success rate, underscoring the potential risks posed by this category of attacks.

%% file: 08_appendix.tex
\clearpage
\setcounter{secnumdepth}{2}
\renewcommand{\thesubsection}{\thesection.\arabic{subsection}}
\appendix
\section{Target Class Mapping Algorithm}\label{algorithm}

Algorithm~\ref{alg:greedy-search} describes the details of our proposed A2X attack.

\begin{algorithm}[ht]
\caption{Mapping Selection}
\label{alg:greedy-search}
\begin{algorithmic}[1]
\Require clean dataset $\mathcal{D}=\{x,y\}$, trigger $t$, mask $m$, training epoch $N$
\Ensure mapping $\mathcal{G}$
\State Train a surrogate model $f_s$ on $\mathcal{D}$
\State Build a poison dataset $\mathcal{D'}=\{(x_{tr},y)\}, x_{tr} = (1 - {m}) \odot {x} + {m} \odot {t} $
\For{$i := 0$ to $K-1$}
    \State $positions[i] \gets \text{calculate position vector on class $i$ of $\mathcal{D'}$ }$
\EndFor
\For{$i,j:= 0$ to $K-1$}
    \State $d[i][j] \gets \| positions[i] -positions[j]  \|_2$
    \State // Let $d[i][j]$ denote the distance between class $i$ and class $j$
\EndFor
\State $\mathcal{C} \gets Kmeans(positions,X)$
\State // Let $\mathcal{C}$ denote the set that includes all groups
\For{$C_i$ in $\mathcal{C}$}
    \For{$j$ in $C_i$}
        \For{$k := 0$ to $K-1$}
            \State groups distance matrix $D[i][k] \gets D[i][k]+d[j][k]$    
            \State // Let $D[i][k]$ denote the distance between group $C_i$ and class $k$
        \EndFor    
    \EndFor
\EndFor
\State $\mathcal{G} \gets Hungarian Algorithm(D)$
\State \Return $\mathcal{G}$
\end{algorithmic}
\end{algorithm}
\section{Comparison Between the A2X Attack and O2N Attack}\label{sec:Why A2X}
We compare the A2X attack with another multi-target backdoor attack, O2N~\cite{xue2020one}, to demonstrate that not all multi-target attacks possess inherent robustness characteristics against defensive methods.
Following the original paper’s setup, we used patches of varying intensities to attack different classes. Experiments were conducted on CIFAR10 with ResNet18, using a 5\% poisoning rate, targeting 2, 5, 8, and 10 classes respectively. We evaluated with the IBD~\cite{hou2024ibd} defense, which performs well against A2O attacks, to test both types of attacks.

\paragraph{Experimental Results:} Table~\ref{tab:O2N} presents the comparison of attack success rates after defense, demonstrating the superior robustness of A2X against defenses. The IBD method effectively defends against O2N, with O2N’s ASR below 40\% across all four settings, while A2X only falls below 40\% when X=2 (1 out of 4).
This is because O2N can be viewed as a simple combination of multiple A2O attacks, where the characteristic vulnerability of A2O attacks---overfitting to the trigger---remains present in O2N. Consequently, many defenses designed for A2O attacks (which constitute the majority of existing defenses) remain effective against O2N. In contrast, A2X attacks break the fundamental assumptions underlying existing defenses and can therefore bypass them.

\begin{table}[htbp]
\centering
\fontsize{9}{9}\selectfont
\setlength{\tabcolsep}{2pt}
\begin{tabular}{c|c|c}
\toprule
\textbf{Defenses}$\rightarrow$ & \textbf{A2X} & \textbf{O2N} \\
\textbf{Value of X}$\downarrow$ & ASR  & ASR    \\
\midrule
\textbf{X=2}   & 5.9\scalebox{0.8}{$\pm$8.6} & 5.7\scalebox{0.8}{$\pm$5.8}  \\
\textbf{X=5}   & 43.0\scalebox{0.8}{$\pm$35.6} & 13.9\scalebox{0.8}{$\pm$13.7}  \\

\textbf{X=8}   & 51.6\scalebox{0.8}{$\pm$34.0} & 34.4\scalebox{0.8}{$\pm$9.9}   \\

\textbf{X=10}  & 72.5\scalebox{0.8}{$\pm$9.1} & 14.0\scalebox{0.8}{$\pm$7.6}\\

\bottomrule
\end{tabular}
\caption{Performance of IBD Defense Methods Against A2X Attacks and O2N Attacks on CIFAR10 with ResNet18.
}
\label{tab:O2N}
\end{table}

\section{More Details of Related Backdoor Defenses} \label{sec:backdoor-defense-appendix}
We categorize existing defense methods into mainly three categories. \textbf{(1) Dataset purification.} The primary goal of these methods is to detect and remove poisoned samples from the training dataset. Li et al.~\cite{li2021anti} observed that poisoned samples converge faster in the early training phase compared to clean ones and leveraged this difference in loss values to separate them. A lot of following works \cite{tang2023setting,zhu2023victim,gao2023backdoor} are all based on similar observations.  Huang et al.~\cite{huang2022backdoor} utilized unsupervised learning to remove anomalous poisoned samples. Chen~\cite{chen2022effective} leveraged semi-supervised learning to mitigate the accuracy degradation in prior unsupervised learning. Pu et al.~\cite{pu2024mellivora} found that poisoned samples are more robust to adversarial perturbations. \textbf{(2) Sample detection.} Gao et al.~\cite{gao2019strip} observed that poisoned samples tend to maintain their original predicted labels even after being mixed with samples from other classes, whereas clean samples usually change their predictions under the same conditions. Similarly, these works \cite{guo2023scale,hou2024ibd} observed that when pixel values of the inputs or the Batch Normalization (BN) parameters of the model are amplified, poisoned samples still persist in outputting their original labels. \textbf{(3) Model detection.} Xu et al.~\cite{xu2021detecting} proposed training a meta-classifier to determine whether a model is backdoored. Wang et al.~\cite{wang2019neural} employed gradient descent to search for reverse triggers. Based on the observation that the trigger of target class tends to be smaller, they applied anomaly detection to determine whether the model contains a backdoor. Ma et al.\cite{ma2024need} utilized the average and difference of images to accelerate the generation of reverse triggers. Based on our analysis, we find that most existing defense methods focus on  A2O attacks,  while the more general multi-target A2X attacks are rarely discussed. As a result, many existing defense methods are not applicable in such scenarios.  

\section{Details of Experimental Setup and Implementation}
\label{experimental settings}
\subsection{Training Settings}
To facilitate the experiments, both the victim and surrogate models adopt the ResNet18 architecture and are trained using the same procedure by default. We use the SGD optimizer with a batch size of 128, an initial learning rate of 0.1, and a total of 100 training epochs. A weight decay of is 0.0005 applied, and the learning rate is reduced by a factor of 10 at the 50th and 80th epochs. For experiments using the Adam optimizer \cite{kingma2014adam}, the initial learning rate is set to 0.005. All experiments are conducted on a machine equipped with a 13th Gen Intel(R) Core(TM) i9-13900KF CPU, 128 GB RAM, 4 TB ROM, and an NVIDIA RTX 4090 GPU.

\subsection{Implementation of Attacks}
\label{detail attack}
Since these attacks are limited to A2O or A2A settings, we adopt only their trigger patterns rather than strictly following the full attack implementations as described in the original papers. The details of backdoor triggers are summarized in Table~\ref{tab:trigger_summary}. 
\begin{table*}[h]

\centering
\begin{tabular}{llll}
\toprule
\textbf{Attacks} & \textbf{Trigger Type} & \textbf{Trigger Pattern} & \textbf{Other Details} \\
\midrule
BD-white & Replace Pattern & White Squre & \\
BD-random & Replace Pattern & Random Squre & \\
Blend & Blend Pattern & Random Pixel & $\alpha = 0.2$ \\
LC & Replace Pattern & Four-corner  \\
SIG & Blend Pattern & Sinusoidal Signal & $\Delta =20 ~f=6$\\

\bottomrule
\end{tabular}
\caption{Summary of Different Trigger Patterns}
\label{tab:trigger_summary}
\end{table*}

\subsection{Implementation of Defenses}
\label{detail defense}
We use six defensive methods in our experiments, including Anti-Backdoor Learning (ABL)
~\cite{li2021anti}, the Victim and the Beneficiary (V\&B)~\cite{zhu2023victim}, FinePruning(FP)~\cite{liu2018fine}, Input-level Backdoor Detection (IBD)~\cite{hou2024ibd}, SCALE-UP(SCALE)~\cite{guo2023scale} and Neural Cleanse (NC)~\cite{wang2019neural}. The implementation details of them are provided as follows:
\begin{itemize}

\item ABL: We find that the learning rate in the unlearning stage of the official code is too high, which caused a significant drop in model accuracy under the clean (non-attacked) setting. Therefore, we reduced the learning rate in the unlearning stage to 0.0001, while keeping all other parameters the same as in the official code.
\item V\&B: V\&B is not included in backdoorbox\cite{li2023backdoorbox}, so we directly used the official code, 3 epochs of warm-up, 100 epochs of supervised learning, and 50 epochs of semi-supervised learning, while keeping all other parameters the same as in the official code..

\item IBD: IBD is a defense for detect posion samples during the inference phase. For ease of comparison, we use this method to filter 1\% of the training samples for unlearning, set the same as ABL. 
\item SCALE: Same as IBD, we use this method to filter 1\% of the training samples for unlearning.
\item FP: We use 10000 reserved clean samples to prune the 20\% of the channels of the second layer.

\item NC: We train the reverse trigger for each class over 50 epochs, and in alignment with the configuration in the original paper, we deploy the unlearning process to facilitate comparison. Distinct from the unlearning strategy employed in ABL, the approach here involves embedding reverse triggers into a subset of clean samples, this method adds reverse triggers into a portion of clean samples and performs fine-tuning jointly with the unmodified clean sample.
\end{itemize}

\section{Detailed Analysis of the Ineffectiveness of Existing Defenses}
Since both ABL and V\&B are based on the assumption that backdoor samples converge faster, while IBD and SCALE rely on the model's overfitting to triggers. Accordingly, redundant discussion on V\&B and SCALE is avoided here.
\label{defense analysis}
\subsection{Analysis of Neural Cleanse (NC)}
\begin{figure}
  \centering
  \includegraphics[width=1\linewidth]{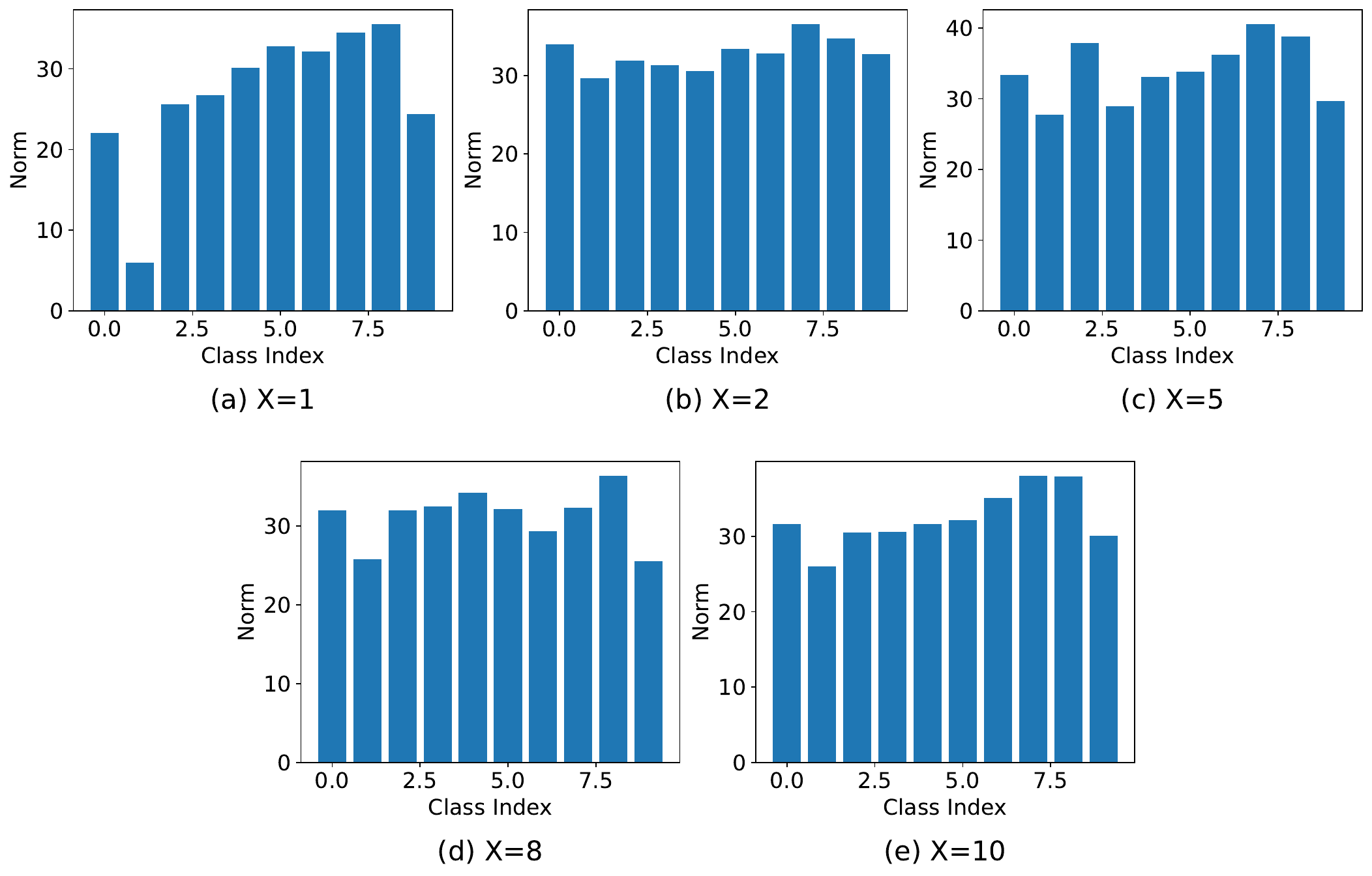}
  \caption{Norm of Reversed Trigger under Different Value of X.}
  \label{fig:9}
\end{figure}

NC assumes that misclassifying samples into the target class requires only minimal perturbation. Therefore, by performing anomaly detection on reverse-engineered triggers for all classes, it is possible to identify the target class. We conduct a statistical analysis of the norm values of the generated triggers under different values of $X$, with the results shown in Figure~\ref{fig:9}. It is evident that when $X=1$, the trigger norm for the target class (class 1) is significantly lower than those of the other classes. In this case, the trigger is successfully identified, and the backdoor is effectively removed. However, in other attack settings, the norm values across all classes are relatively similar, without any clear outliers. As a result, NC mistakenly concludes that the model is clean and fails to detect the backdoor.

\subsection{Analysis of Anti-Backdoor Learning (ABL)}
\begin{figure}
  \centering
  \includegraphics[width=1\linewidth]{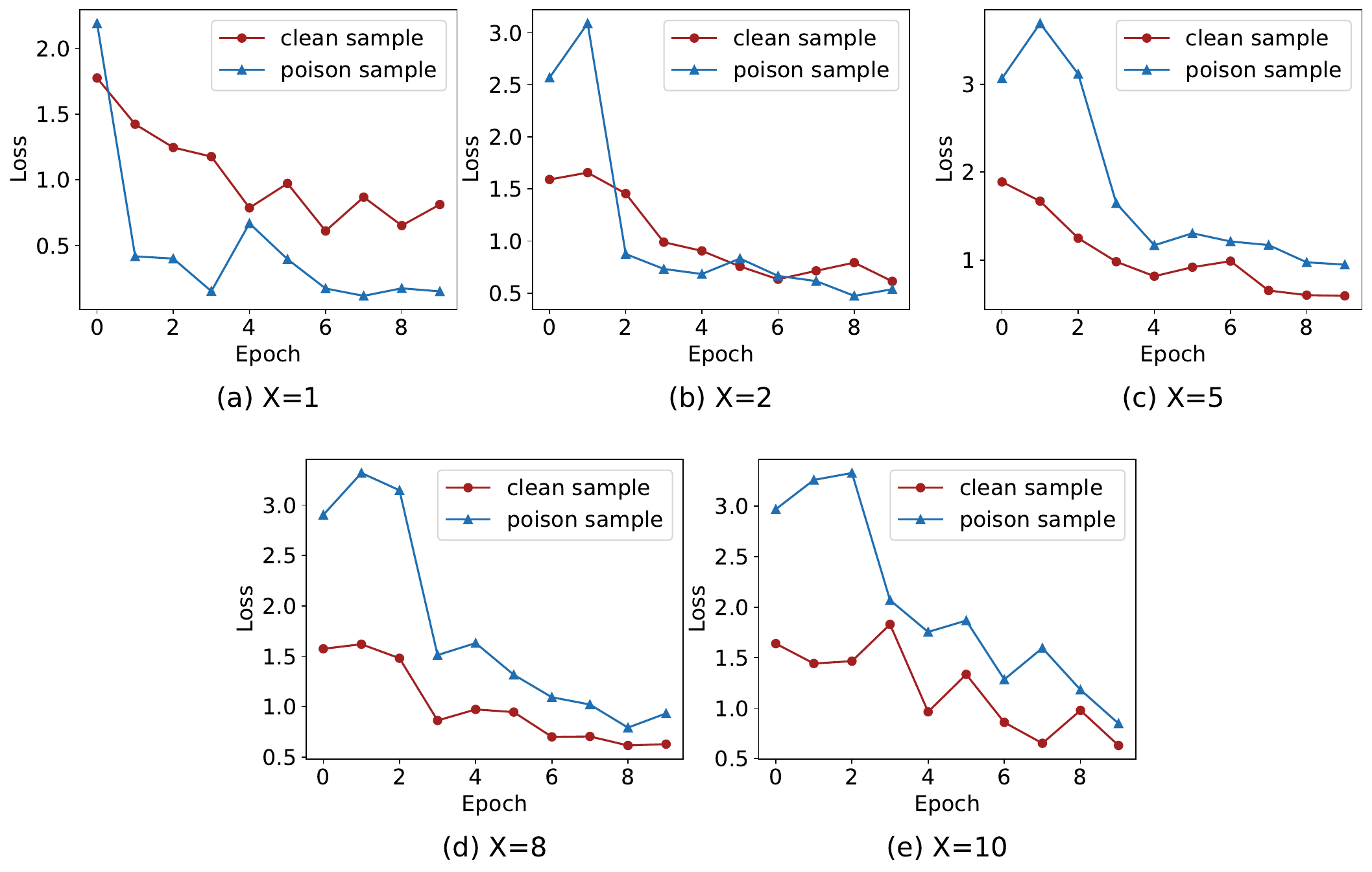}
  \caption{Training Loss Trends for Clean vs. Poisoned Samples (Early Epochs).}
  \label{fig:10}
\end{figure}

ABL hypothesizes that poisoned samples converge faster than clean samples during the early stages of training. Figure~\ref{fig:10} shows the changes in the loss  of poisoning and clean samples under different values of $X$. It can be observed that the poisoned samples exhibit a rapid loss drop only when $X=1$, the hypothesis of ABL gradually fails as $X$ increases. At $X=10$, the loss of poisoned samples even exceeds that of clean ones. These observations suggest that relying on loss discrepancy for backdoor defense is unreliable, as attackers can easily manipulate X to mimic the convergence behavior of clean samples. 

\subsection{Analysis of Input-level Backdoor Detection (IBD)}
\begin{figure}
  \centering
  \includegraphics[width=1\linewidth]{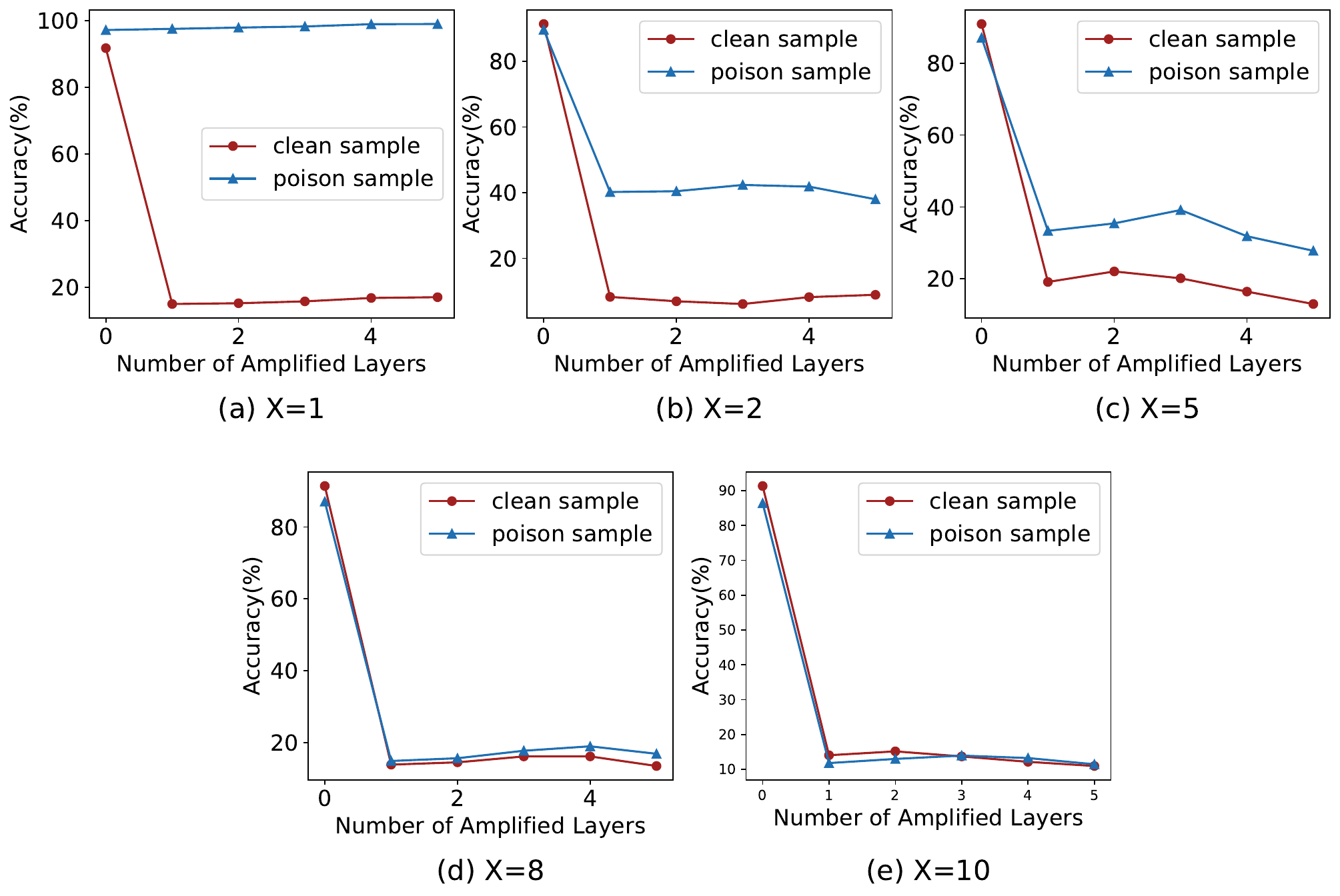}
  \caption{The Accuracy of clean and Poisoned Samples when Amplifying Different Numbers of BNlayers.}
  \label{fig:11}
\end{figure}

IBD observes that due to the overfitting to the trigger, the prediction of poisoned samples remains stable when BatchNorm (BN) layer parameters are amplified, while the prediction of clean samples is highly sensitive to such amplification. Similarly, under different values of $X$, we evaluate the changes in accuracy as the number of amplified BN layers increases, as shown in Figure~\ref{fig:11}. A stable accuracy of poisoned samples is only observed when $X=1$, while in other cases, the accuracy drops rapidly. This suggest that the degree of overfitting decreases with increasing $X$. Therefore, this defense strategy is also invalid.

\subsection{Analysis of Fine Pruning (FP)}
FP is based on the distinction between backdoor and clean sample recognition mechanisms, assuming the existence of neurons highly correlated with the backdoor task. By pruning these neurons, effective defense can be achieved. However, this assumption becomes invalid in scenarios where $X>1$, as backdoor samples must first recognize their original class, this process highly similar to that of clean samples. As a result, such methods struggle to remove backdoors effectively.

\section{More Details of Experimental Results}
\begin{figure*}[htbp]
  \centering
  \includegraphics[width=0.9\linewidth]{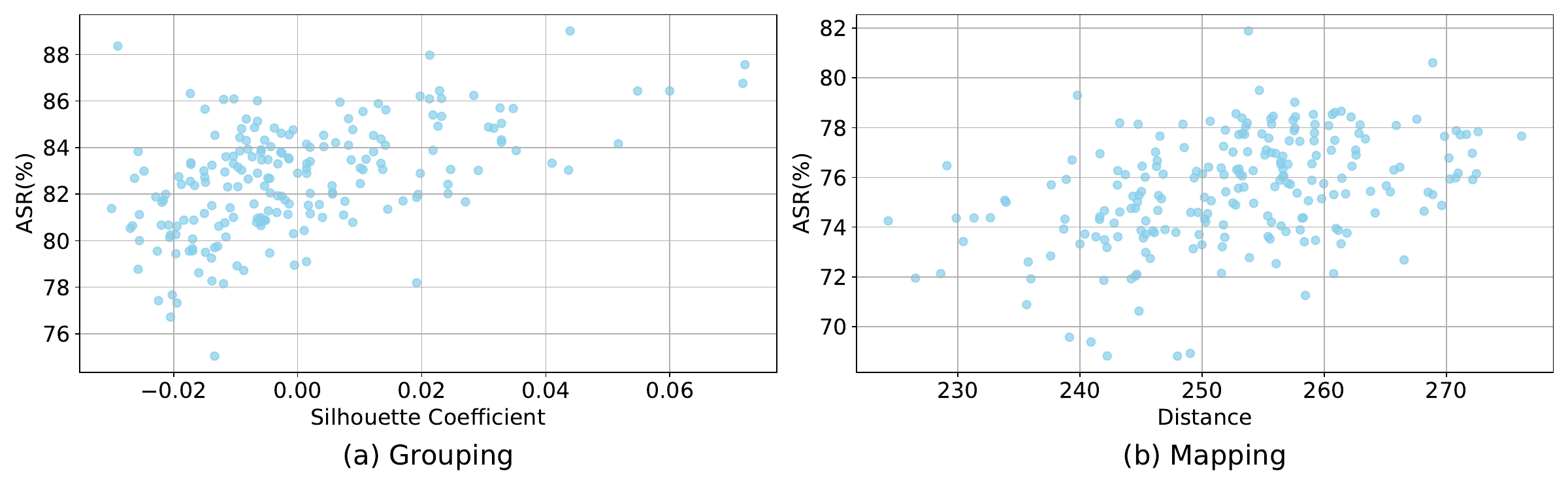}
  \caption{The Attack Success Rate under Different Silhouette Coefficient and Distance Between Groups and Target Classes.}
  \label{fig:8}
\end{figure*}
\subsection{Empirical Validation of Similarity-based Class Grouping}
\label{empirical_vaildation_grouping}
Based on the grouping strategy in Section~\ref{motivation}, we hypothesize that merging similar classes into the same group can lead to more effective backdoor attacks. To verify this hypothesis, we conduct the following experiment. According to the results shown in Figure~\ref{fig:ablation_studies}, in order to eliminate the influence of  target class assignment strategies, we choose a smaller value of $X$. Specifically, we conduct an A2X($X=2$) attack with a poisoning rate of 0.5\% on the CIFAR10 dataset. To evaluate the quality of the grouping, we adopt the commonly used average silhouette coefficient $\bar{s}$, which is defined as follows:

\begin{equation}
\label{silhouette coefficient}
  \bar{s} \;=\; \frac{1}{K} \sum_{i=1}^{K} s(i)\;=\;\frac{1}{K}\sum_{i=1}^{K}\frac{b(i) - a(i)}{\max\{a(i),\,b(i)\}}
\end{equation}
$s(i)$ denotes the silhouette coefficient of the $i$-th class, $a(i)$ represents the average distance between class $i$ and the other classes within the same group, $b(i)$ denotes the average distance between class $i$ and the classes in the nearest neighboring group. The value of $s(i)$ lies in the range [-1,1], where a higher value indicates a better quality of the grouping. To evaluate the impact of silhouette coefficient, we generate 200 random mappings, compute their silhouette coefficients, and use them to train models for backdoor attacks. As illustrated in the Figure~\ref{fig:8}(a), we observe the following: (1) Overall, the silhouette coefficient exhibits a positive correlation with the attack success rate, with a Pearson Correlation Coefficient of 0.514. (2) Under random grouping, most mappings yield low silhouette coefficients, with only a few resulting in relatively high values. This explains the generally poor performance of random grouping strategy in terms of attack success rate.

\subsection{Empirical Validation of Distance-aware Target Class Assignment}
\label{empirical_vaildation_mapping}
In the distance-aware target class assignment strategy of Section~\ref{motivation}, we hypothesize that for a given mapping $\mathcal{G}$, as the distance between the groups and the target classes increases, the attack success rate will also increase. In this section, we aim to validate this hypothesis through the following experiment. Based on the conclusion from Figure~\ref{fig:ablation_studies}, we know that when X equals the number of classes, class grouping loses its effectiveness, which helps us eliminate the interference from class grouping. Therefore, we conduct a 0.5\% A2A attack on CIFAR10, and the results are shown in Figure~\ref{fig:8}(b). It can be observed that there is a positive correlation between attack success rate and distance. The Pearson correlation coefficient, calculated to be 0.416, indicates a positive correlation between the two. However, since the training process is influenced by many other factors, this correlation is not statistically significant. Nonetheless, this supports the validity of our hypothesis.

\subsection{Different Types of Triggers}
\label{sec:other trigger}
\begin{figure*}[h]

\centering
\includegraphics[width=0.9\linewidth]{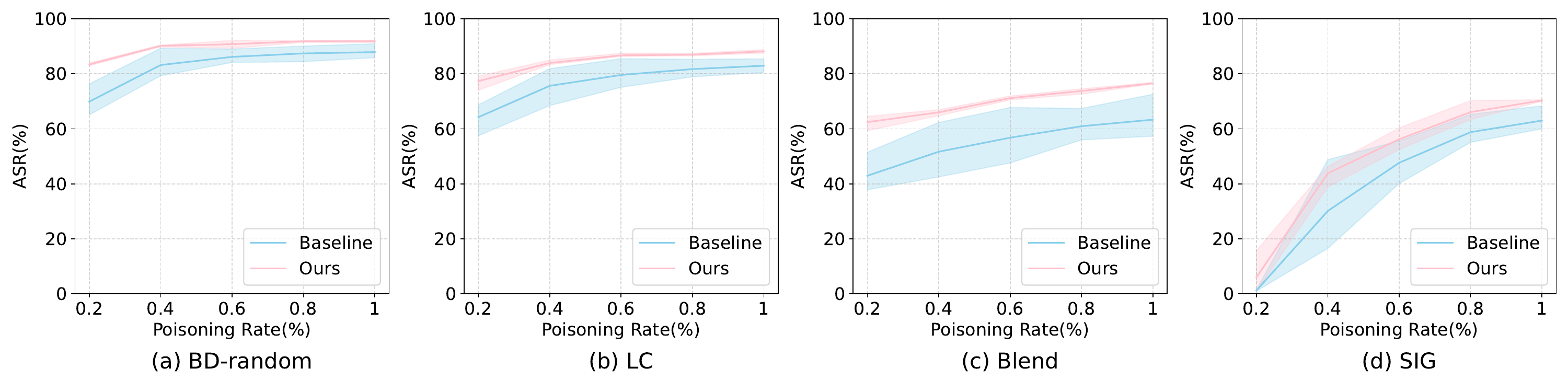}
\caption{The Attack Success Rate of Our A2X attack and Baseline under Four Different Triggers.}
\label{other trigger}
\end{figure*}

We conducted A2X($X=5$) attack using different types of triggers on CIFAR10. As shown in Figure~\ref{other trigger},our method achieves significant improvements across all four triggers, demonstrating its general applicability to arbitrary triggers. Additionally, we observed that replace patterns perform higher attack success rates compared to blend patterns. Specifically, the attacks based on replace pattern, BD-white, BD-random, and LC achieve approximately 90\% attack success rate under a poisoning ratio of just 0.1\%, whereas the attacks based on blend pattern both achieve less than 80\% under the same setting. This is because blend pattern triggers are applied over the entire image, which can interfere with the recognition of the clean features. Since A2X attacks rely on the clean features, the backdoor task becomes more difficult to learn. In contrast, replace pattern triggers are localized and introduce minimal interference to the clean features. Overall, localized triggers that differ more from the clean features tend to be more effective.

\subsection{The Impact of Our A2X Aattack on Accuracy}
\label{ACC}
\begin{table}[t]
\centering
\setlength{\tabcolsep}{2pt}
\fontsize{9}{9}\selectfont
\begin{tabular}{cc|cccc}
\toprule

\multicolumn{2}{c|}{\textbf{Rate}$\downarrow$} & \textbf{X=2} & \textbf{X=5} & \textbf{X=10} \\
\midrule
\multicolumn{2}{c|}{\textbf{0.2\%}} & -0.2\scalebox{0.8}{$\pm$0.2}
& -0.3\scalebox{0.8}{$\pm$0.1} & 0.1\scalebox{0.8}{$\pm$0.1}

\\
\midrule
\multicolumn{2}{c|}{\textbf{0.4\%}} & -0.3\scalebox{0.8}{$\pm$0.3}
& -0.4\scalebox{0.8}{$\pm$0.2} & -0.1\scalebox{0.8}{$\pm$0.1}

\\
\midrule
\multicolumn{2}{c|}{\textbf{0.6\%}} & 0.1\scalebox{0.8}{$\pm$0.1}
& -0.1\scalebox{0.8}{$\pm$0.2} & -0.1\scalebox{0.8}{$\pm$0.3}

\\
\midrule
\multicolumn{2}{c|}{\textbf{0.8\%}} & 0.0\scalebox{0.8}{$\pm$0.2}
& -0.2\scalebox{0.8}{$\pm$0.2} & -0.1\scalebox{0.8}{$\pm$0.2}

\\
\midrule
\multicolumn{2}{c|}{\textbf{1\%}} & -0.2\scalebox{0.8}{$\pm$0.2}
&0.0\scalebox{0.8}{$\pm$0.3}  & 0.0\scalebox{0.8}{$\pm$0.1}

\\

\bottomrule
\end{tabular}
\caption{The Accuracy Difference (\%) between the Backdoored Model and the Clean Model under Different Poisoning Rates in Our A2X Attack.}
\label{tab:acc}

\end{table}
To demonstrate that our proposed A2X attack does not affect the accuracy of the original task, we evaluate the accuracy difference between the backdoored model and the clean model on the CIFAR10 and ResNet18 (the clean model’s accuracy is 91.7\%) under different poisoning rates and different values of target classes (x). Table~\ref{tab:acc} shows the results. Under all settings, the accuracy difference between the backdoored model and the clean model remains below 0.5\%, demonstrating that our A2X attack has minimal impact on the original task and exhibits strong stealthiness.

\subsection{Ablation Studies}
\label{sec:ablation}

\begin{figure}[h]
\centering
\includegraphics[width=0.9\columnwidth]{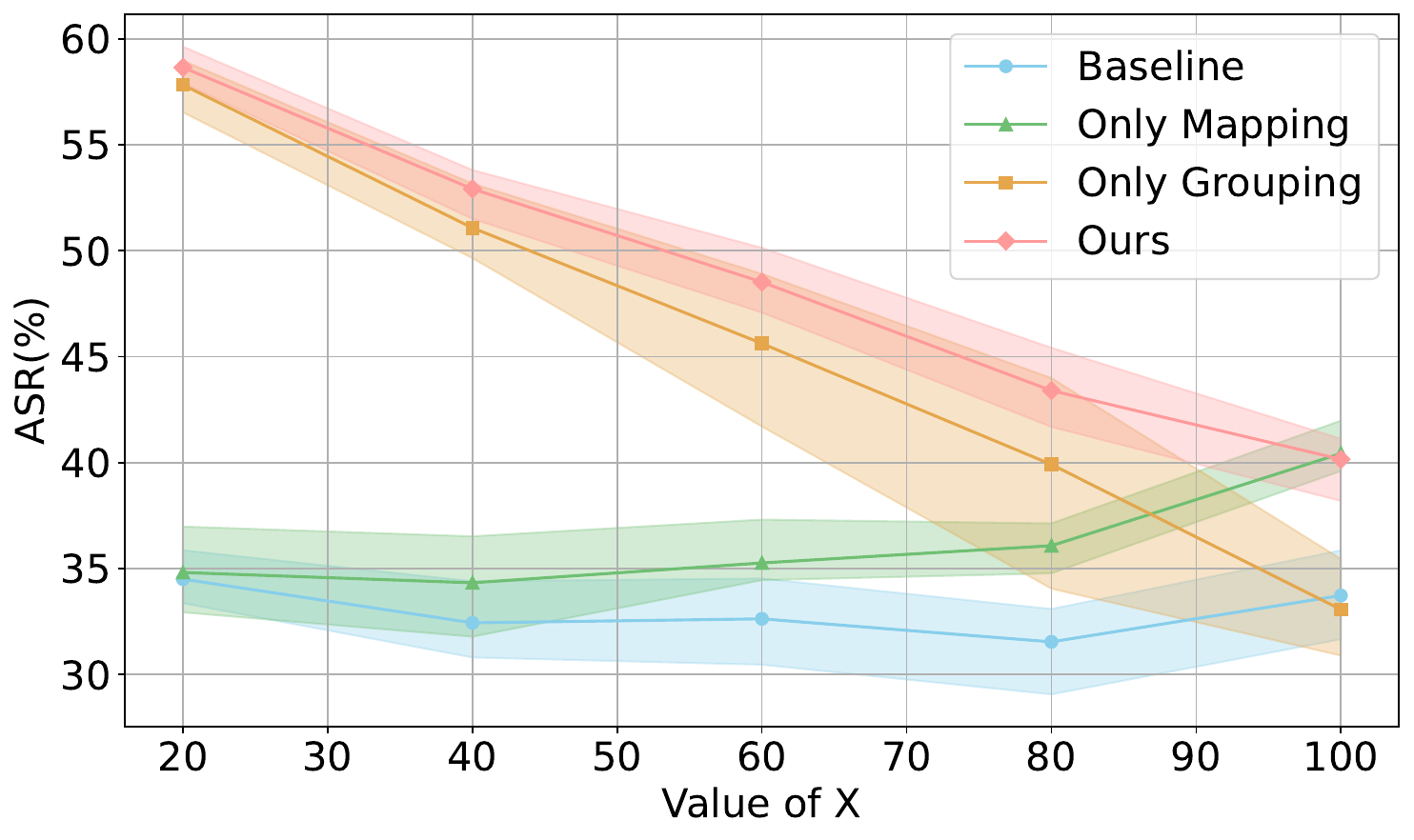} 
\caption{Attack Success Rates across Ablation Configurations with Varying X on CIFAR100 and ResNet18. ``Only Mapping'' lines are the results of only using distance-aware target class assignment, ``Only Grouping'' lines show the results of only using  similarity-based class grouping.}
\label{fig:ablation_studies}
\end{figure}
Our proposed method consists of two main components: similarity-based class grouping and distance-aware target class assignment. We conducted experiments to evaluate the impact of each component on attack effectiveness. We used the attack setup from the CIFAR100 settings in Section~\ref{sec:main_results} with a poisoning rate of 2.5\%, while selectively employing different combinations of our proposed components.

Specifically, in addition to the baseline method and our full method, we designed two intermediate configurations for our ablation studies. The first configuration, called ``only mapping'', replaces our similarity-based class grouping with random grouping while retaining our distance-aware target class assignment. The second configuration, called ``only grouping'', employs our similarity-based class grouping but uses random target class assignment instead of our distance-aware approach.

\textbf{Experimental Results:} Figure~\ref{fig:ablation_studies} shows our results. Removing either the class grouping or target class assignment component leads to decreased attack effectiveness, highlighting the importance of each component in our proposed attack. For $X$=20, the attack success rate drops by 23.84\% and 0.83\% for the ``only mapping'' and ``only grouping'' approaches, respectively. This performance gap widens for ``only grouping'' as $X$ increases, resulting in more significant attack success rate drops (7.09\% when $X$=100). The gap narrows for ``only mapping'' as $X$ increases, but still demonstrates superior performance of the full method on all settings except the case $X$=100 (where the difference of the two methods is less than 1\%).

While both intermediate approaches achieve lower attack success rates compared to our full method that incorporates all design components, it is notable that both still outperform the baseline method that relies entirely on random strategies. The ``only grouping'' approach achieves performance gains similar but slightly lower than our full method--the ``only grouping'' and ``ours'' lines in Figure~\ref{fig:ablation_studies} appear close to each other. The ``only mapping'' approach achieves a relatively smaller improvement in attack success rate, but the enhancement remains significant, exceeding 7\% when $X \geq 80$.

\subsection{Different Epochs of Surrogate Model}

\begin{figure}[h]
    \centering
    \includegraphics[width=0.9\columnwidth]{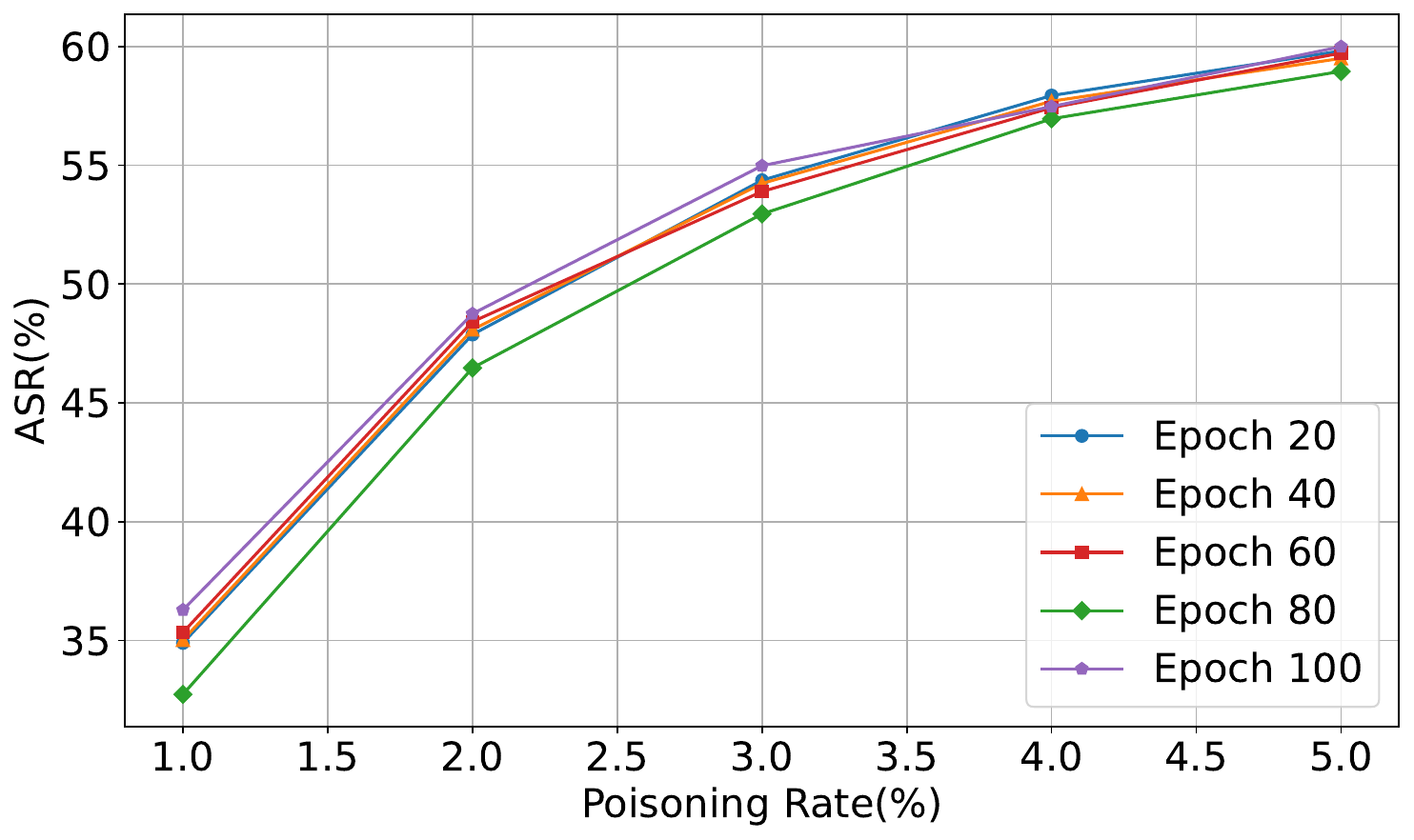} 
    \caption{The Attack Success Rate of Our A2X Method under Five Different Surrogate Model Training Epochs.}
    \label{fig:6}
\end{figure}

We evaluate the impact of different training epochs of the surrogate model, the results are shown in Figure~\ref{fig:6}. Specifically, we consider five increasing epoch values, 20, 40, 60, 80, 100. To more clearly observe the performance differences, we conduct A2X($X=50$) attack on CIFAR100. The results suggest that the number of training epochs has a slight influence on the attack success rate. Among all settings, training for 100 epochs achieves the best performance, which is also our default parameter. In general, the attack success rate tends to increase with more training epochs. This is because a larger number of epochs allows the surrogate model to better capture the features of the dataset, leading to more effective mapping selection. However, considering the trade-off between performance improvement and time cost, and given that the improvement beyond a certain point is relatively marginal, we ultimately select 100 epochs as the default parameter.
\begin{table*}[h]
\centering
\setlength{\tabcolsep}{5pt}
\begin{tabular}{l|cc|cc|cc|cc|cc|cc|cc|cc|cc}
\toprule
\textbf{Values of X}$\rightarrow$ &   \multicolumn{2}{c|}{\textbf{X=2}} & \multicolumn{2}{c|}{\textbf{X=3}} & \multicolumn{2}{c|}{\textbf{X=4}} & \multicolumn{2}{c}{\textbf{X=5}}&\multicolumn{2}{c|}{\textbf{X=6}} & \multicolumn{2}{c|}{\textbf{X=7}} & \multicolumn{2}{c|}{\textbf{X=8}} & \multicolumn{2}{c}{\textbf{X=9}}& \multicolumn{2}{c}{\textbf{X=10}} \\
 & S & T & S & T & S & T & S & T  & S & T & S & T & S & T & S & T & S & T \\
\midrule
\multirow{10}{*}{\textbf{Labels}}    
 & 0 & 5 & 0 & 6 & 0 & 5 & 0 & 5 &0 & 5 & 0 & 5 & 0 & 5 & 0 & 5 & 0 & 5 \\
 & 1 & 5 & 1 & 4 & 1 & 5 & 1 & 5 &1 & 4 & 1 & 2 & 1 & 4 & 1 & 2 & 1 & 4  \\
 & 2 & 5 & 2 & 4 & 2 & 9 & 2 & 1 &2 & 1 & 2 & 1 & 2 & 9 & 2 & 9 & 2 & 9 \\
 & 3 & 9 & 3 & 1 & 3 & 0 & 3 & 9 &3 & 9 & 3 & 9 & 3 & 8 & 3 & 8 & 3 & 8\\
 & 4 & 5 & 4 & 6 & 4 & 1 & 4 & 1 &4 & 1 & 4 & 1 & 4 & 1 & 4 & 1 & 4 & 1\\
 & 5 & 9 & 5 & 1 & 5 & 0 & 5 & 9&5 & 9 & 5 & 9 & 5 & 0 & 5 & 0 & 5 & 0 \\
& 6 & 9 & 6 & 4 & 6 & 9 & 6 & 0 &6 & 0 & 6 & 0 & 6 & 7 & 6 & 7 & 6 & 7 \\
& 7 & 5 & 7 & 6 & 7 & 1 & 7 & 6 &7 & 6 & 7 & 6 & 7 & 6 & 7 & 6 & 7 & 6\\
& 8 & 5 & 8 & 4 & 8 & 5 & 8 & 5 &8 & 4 & 8 & 4 & 8 & 4 & 8 & 4 & 8 & 3\\
& 9 & 5 & 9 & 6 & 9 & 5 & 9 & 5 &9 & 4 & 9 & 2 & 9 & 4 & 9 & 2 & 9 & 2\\
\bottomrule
\end{tabular}
\caption{Examples of Optimal Class Mappings Identified by Our Method on CIFAR10, S and T denote as Source Class ad Target Class. }
\label{tab:example1}
\end{table*}
\subsection{Different Distance Norms} \label{sec:discussion_of_norms}
We used the attack setup from the CIFAR10 settings with a poisoning rate of 0.5\% to evaluate the impact of different distance norms, $\ell_1$,$\ell_2$,$\ell_\infty$. According to the results reported in Table~\ref{tab:norm}, under different values of X,  the results across different distance norms exhibit only slight differences, about 1\%-2\%, but all consistently outperform the baseline. Among the three, the best improvement is achieved by our default choice, $\ell_2$ norm, which obtains the highest attack success rate in all cases except when $X=2$. In addition, the $\ell_2$ norm yields the lowest standard deviation, indicating that the corresponding backdoor attack is also more stable. Overall, the $\ell_2$ norm achieves the best trade-off between attack effectiveness and stability.

\begin{table}[htbp]

\small
\centering
\setlength{\tabcolsep}{2pt}
\begin{tabular}{c|c|c|c|c|c}
\toprule
\textbf{Value of X}$\rightarrow$ & \textbf{X=2} & \textbf{X=4} & \textbf{X=6} &\textbf{X=8} & \textbf{X=10} \\
\midrule  
\textbf{$\ell_1$} & \textbf{85.5}\scalebox{0.8}{$\pm$ 1.4} & 84.0\scalebox{0.8}{$\pm$ 0.9} & 81.4\scalebox{0.8}{$\pm$ 1.0} & 79.7\scalebox{0.8}{$\pm$ 1.4} &77.4\scalebox{0.8}{$\pm$ 1.5} \\
\textbf{$\ell_2$} & 84.5\scalebox{0.8}{$\pm$ 1.8} & \textbf{84.5}\scalebox{0.8}{$\pm$ 0.6} & \textbf{83.2}\scalebox{0.8}{$\pm$ 0.9} & \textbf{80.3}\scalebox{0.8}{$\pm$ 1.1} & \textbf{78.8}\scalebox{0.8}{$\pm$ 1.3} \\
\textbf{$\ell_\infty$} & 85.1\scalebox{0.8}{$\pm$ 0.9} & 83.5\scalebox{0.8}{$\pm$ 1.4} & 81.5\scalebox{0.8}{$\pm$ 1.4} & 78.4\scalebox{0.8}{$\pm$ 1.8} &77.8\scalebox{0.8}{$\pm$ 1.5} \\
\textbf{$baseline$} & 81.7\scalebox{0.8}{$\pm$ 2.7} & 78.9\scalebox{0.8}{$\pm$ 1.6} & 74.8\scalebox{0.8}{$\pm$ 3.0} & 74.7\scalebox{0.8}{$\pm$ 1.9} &75.9\scalebox{0.8}{$\pm$ 1.9} \\

\bottomrule
\end{tabular}
\caption{ The Attack Success Rate(\%) of Our A2X Attack under Different Distance Norms}
\label{tab:norm}
\end{table}

\subsection{Examples of Mappings}

Table~\ref{tab:example1} presents the optimal class mappings identified by our method on the CIFAR10 dataset under different values of $X$. The class labels are: 0 – airplane, 1 – automobile, 2 – bird, 3 – cat, 4 – deer, 5 – dog, 6 – frog, 7 – horse, 8 – ship, and 9 – truck.

\section{The impact of mapping strategy on attack flexibility}
Our A2X attack can be easily adapted to accommodate more flexible mapping choices. The attackers can first fix the required mappings and then use our method to generate the remaining ones. This work primarily highlights the threats posed by A2X attacks, while the attack flexibility will be further studied in future work.
